\documentclass{article}
\usepackage[utf8]{inputenc}
\usepackage{graphicx}
\usepackage{subcaption}
\usepackage{amsmath}
\usepackage{tablefootnote}
\usepackage{csvsimple}
\usepackage{setspace}
\usepackage{amsthm}
\usepackage[ruled,vlined]{algorithm2e}
\usepackage{tikz}
\usepackage{blkarray}
\usepackage{hyperref}
\usepackage[section]{placeins}
\usetikzlibrary{arrows}
\usepackage[
  backend=biber,
  style=apa,
  citestyle=apa
]{biblatex}
\usepackage{geometry}
\newtheorem{definition}{Definition}

\newtheorem{corollary}{Corollary}
\newtheorem{theorem}{Theorem}
\def\indep{\perp \!\!\! \perp}

\title{Causal Discovery of Macroeconomic State-Space Models\thanks{Contact: \href{mailto:emmet.hall-hoffarth@economics.ox.ac.uk}{emmet.hall-hoffarth@economics.ox.ac.uk}. I would like to thank Jeremy Large and Michael McMahon for their thoughtful comments and guidance in the writing of this paper.}}
\author{Emmet Hall-Hoffarth\\University of Oxford}
\date{\today}

\makeatletter
\newenvironment{chapquote}[2][2em]
  {\setlength{\@tempdima}{#1}%
   \def\chapquote@author{#2}%
   \parshape 1 \@tempdima \dimexpr\textwidth-2\@tempdima\relax%
   \itshape}
  {\par\normalfont\hfill--\ \chapquote@author\hspace*{\@tempdima}\par\bigskip}
\makeatother

\begin{document}

\maketitle

\abstract{This paper presents a set of tests and an algorithm for agnostic, data-driven selection among macroeconomic DSGE models inspired by structure learning methods for DAGs. As the log-linear state-space solution to any DSGE model is also a DAG it is possible to use associated concepts to identify a unique ground-truth state-space model which is compatible with an underlying DGP, based on the conditional independence relationships which are present in that DGP. In order to operationalise search for this ground-truth model, the algorithm tests feasible analogues of these conditional independence criteria against the set of combinatorially possible state-space models over observed variables. This process is consistent in large samples. In small samples the result may not be unique, so conditional independence tests can be combined with likelihood maximisation in order to select a single optimal model. The efficacy of this algorithm is demonstrated for simulated data, and results for real data are also provided and discussed.}

\vspace{1cm}

\pagenumbering{arabic}

\begin{chapquote}{Guido Imbens, \textit{Journal of Economic Literature, 2020}}
  "... the most important issue holding back the DAGs is the lack of convincing empirical applications. History suggests that those are what is driving the adoption of new methodologies in economics and other social sciences, not the mathematical elegance or rhetoric."
\end{chapquote}

\vspace*{\fill}

\noindent{\textbf{JEL}: E17; C63; C68} \\
\noindent{\textbf{Keywords}: Baysian Network, DAG, DSGE, State Variable, Macroeconomics} \\

\newpage

\tableofcontents

\newpage

\section{Introduction}

In the machine-learning literature, causal discovery is generally defined as the act of inferring causal relationships from observational data \parencite{huang2020causal}. This however also exactly describes the goal of much of empirical economic research, and therefore, in this context it is most reasonable to append to this definition that which is taken for granted in machine-learning --- that this inference is done \textit{algorithmically}. The field of (algorithmic) causal discovery has benefited from intense development in recent years, however, it is hardly a new discipline. Work along these lines started in the 1980s with early contributions from Judea Pearl, Thomas Verma, and Peter Spirtes, among others. Indeed, there has been considerable work done in the field of economics regarding algorithmic model selection, in particular the general-to-specific model selection of \citeauthor{krolzig2001computer} (\citeyear{krolzig2001computer}).

While there are many approaches to causal discovery, the current paper focuses on the inference of a Directed Acyclical Graph (DAG), sometimes also (somewhat misleadingly) referred to as a Bayesian Network\footnote{This is somewhat misleading because Bayesian Networks do not require the application of Bayes rule, although one could choose to estimate the associated parameters in this way.}. These are a type of \textit{graphical model} which can be used to illustrate, and in many cases infer, causal relationships between variables. While the use of these models as a descriptive tool has been fiercely debated \parencite{pearl2018book}, what is perhaps more exciting for the field of economics is the fact that numerous algorithms exist which, under relatively mild conditions, can identify a DAG, and thus a causal model, directly from observational data.

While there is a considerable potential for the application of such a tool in economics, thus far relatively little work in this vein has taken place. Indeed, \citeauthor{imbens2019potential} (\citeyear{imbens2019potential}) considers the value of DAGs for empirical economics and concludes that the reason this framework has not caught on is precisely because few useful applications have been demonstrated. Notable exceptions include the work of \citeauthor{demiralp2003searching} (\citeyear{demiralp2003searching}), who consider structure learning algorithms for DAGs in the context of Structural Vector Autoregressions (SVARs), and \citeauthor{bazinas2015causal} (\citeyear{bazinas2015causal}), who utilise concepts of conditional (in)dependence closely related to those used in DAGs to develop the notion of \textit{causal transmission}. Notwithstanding these, this paper aims to provide a substantive contribution to the literature by presenting an application of DAGs to macroeconomic DSGE models. In particular, I show that a DSGE model's log-linear state-space solution can be represented as a DAG, and that the structure of that DAG, and thus that of the state-space solution, can be recovered consistently from observational data alone. 

DSGE models such as the \textit{Real Business Cycle} (RBC) model first popularised by \citeauthor{kydland1982time} (\citeyear{kydland1982time}), and subsequent \textit{New Keynesian} models were formulated primarily as a response to the \textit{Lucas critique}; that reduced form macroeconomic time-series models such as VARs are unsuitable for inferring the causal effects of changes to microeconomic or structural parameters or of truly exogenous (uncorrelated) shocks \parencite{lucas1976econometric}. The key feature of DSGE models is that they are based on \textit{microfoundations} --- that is, they explicitly model the optimal behaviour of economic agents in order to derive equilibrium conditions among observed macroeconomic variables. However, these optimisation problems are still subject to assumptions about the nature of constraints faced by agents, the information available to them, and in some cases even their degree of rationality. For example, do agents form expectations in a purely forward-looking fashion, or do they employ some form of indexing to past values? In the relevant literature these assumptions are generally justified either with microeconomic evidence or comparing the \textit{impulse response functions} generated by the model to those estimated by econometric models \parencite{christiano2018on}. 

Different assumptions about microfoundations will sometimes, but not always, imply different state-space models. For example, in standard DSGE models consumption is a control variable, however, if habits in consumption are assumed it becomes a state variable \parencite{fuhrer2000habit}, such that the past value of consumption becomes relevant in determining the current value of other variables in the model. In these cases, the test and algorithm presented in this paper can be seen as another tool that can be used to evaluate these types of assumptions. This evidence is particularly valuable because it is obtained in a \textit{maximally agnostic} way that makes no assumptions about any of the particular observables (e.g. inflation, interest, output), only about the nature of the relationships between them\footnote{We assume linearity and Gaussian shocks for simplicity, although in principle these too could be relaxed.}. In other words, the algorithm regards any observables as ex-ante equally likely to be either state variables or controls, so any conclusions drawn in this way solely reflect the data to the greatest extent possible. What this paper does not do is present a solution to the problem of \textit{microeconomic dissonance} \parencite{levin2008macroeconometric}. In cases where the linear state-space model implied by DSGE models are equivalent, this procedure cannot determine which set of microfoundations are more reasonable.

In order to test the ability of various algorithms in practice I generate random observations from well known DSGE models and then test the ability of various algorithms to identify the ground-truth, which in this context is known. Despite considerable promise, and theoretical guarantees of asymptotic consistency, in these simulation experiments existing structure learning algorithms for DAGs performed poorly at identifying the correct ground-truth state-space model. This is likely due to the fact that these algorithms search over the set of all possible DAGs, of which those that are also state-space models are only a small subset. This is compounded by the fact that in macroeconomics sample sizes available are usually small relative to the number of observables. Conversely, the algorithm I propose explicitly assumes that the solution is within the subset of DAGs which are also state-space models. It is asymptotically consistent, and thanks to the smaller search space, simulation evidence demonstrates that it is also much more successful than existing structure learning algorithms at identifying the ground-truth state-space model, even given realistic (small) sample sizes.

The remainder of this paper is organised as follows. Section \ref{lit_review} covers background information on both DAGs and DSGE models. Section \ref{methodology} introduces the proposed structure learning tests and algorithm. Section \ref{data} briefly introduces the simulated and real world data which will be used for empirical validation. Section \ref{results} provides and discusses the results of the proposed algorithm, as well as some existing alternatives on those data sets. Section \ref{conclusion} includes some closing remarks.

\section{Literature Review} \label{lit_review}

\subsection{DAGs} \label{dags}

\subsubsection{Preliminaries} \label{prelim}

\begin{figure}
  \centering
  \includegraphics[width=0.75\textwidth]{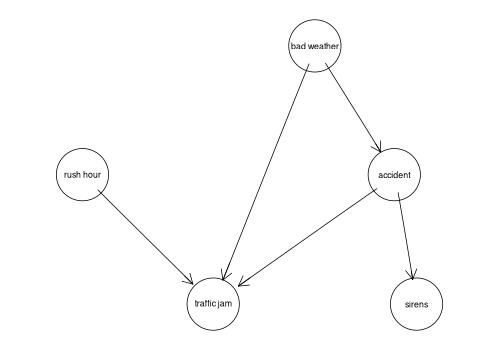}
  \caption{A simple example of a DAG \parencite{traffic_jam}}
  \label{dag1}
\end{figure}

Formally, a DAG $G$ is a pair $(V,E)$ where $V$ is a set of \textit{nodes}, one for each of $k$ observable variables, and $E$ is a $k \times k$ matrix of \textit{edges} or \textit{arcs} \parencite{kalisch2007estimating}. $(x,y) \in$ $E$ indicates the presence of a directed edge from node $x$ to node $y$. As the name DAG suggests, every edge in $E$ is directed such that if $(x,y) \in E$ then $(y,x) \not \in E$. $E$ is also assumed to not contain any cycles, that is, there is no set of edges $\{p_1, p_2, ..., p_k | p_i \in E\}$ containing a directed path starting and ending at the same node. Figure \ref{dag1} gives a simple example of a DAG.

In general, DAGs can represent either discrete, continuous, or mixed variables, but in the current application only continuous variables will be considered. For simplicity, each arc will hereafter be assumed to define a linear relationship between continuous variables. With this assumption we can more specifically define $V$ as a $(k \times 1)$ vector and $E$ as a $k \times k$ adjacency matrix containing slope parameters, where $e_{ij} \not = 0$ indicates a directed edge from node $i$ to node $j$ and $e_{ij} = 0$ indicates the lack of an edge. The directedness assumption is as before, and the acyclic property is now equivalent to the statement that $E^n$ has zeros on its diagonal for $\forall n > 0$. In the spirit of traditional econometric SVARs, the model will now also include a $k \times 1$ vector $\mathbf{\epsilon}$ containing mutually independent Gaussian shocks, one for each node.

The set of nodes from which an arc into a node $x$ originates in a graph $G$ are known as the \textit{parents} of $x$ ($pa_G(x)$), and the set of nodes that have an incoming arc from $x$ are known as the \textit{children} of $x$ ($ch_G(x)$) \parencite{pearl2009causality}. The set of all nodes from which a directed path into $x$ originates are known as the \textit{ancestors} of $x$ ($ans_G(x)$) and the set of all nodes that have an incoming path from $x$ are known as the \textit{decedents} of $x$ ($des_G(x)$). 

I will now briefly review some key results pertaining to DAGs that are leveraged in this paper. For a more complete treatment see \citeauthor{pearl2009causality} (\citeyear{pearl2009causality}).

\theoremstyle{definition}
\begin{definition}{Stability}
  Let $f(\mathbf{w};\theta)$ represent some DGP defined over observable variables $\mathbf{w}$ with true parameters $\theta$, and $I(f(\mathbf{w};\theta))$ be all the conditional independence relationships that exist in $f(\mathbf{w};\theta)$ between variables in $\mathbf{w}$. Then $f$ is \textbf{stable} if $I(f(\mathbf{w};\theta)) = I(f(\mathbf{w};\theta^\prime)) \text{  } \forall \text{  } \theta^\prime$ in a neighbourhood of $\theta$. \parencite[p.48]{pearl2009causality}
  \label{stability}
\end{definition}

\begin{definition}{Faithfulness}
   A DAG $G$ is said to be \textbf{faithful} to $f(\mathbf{w};\theta)$ if $f(\mathbf{w};\theta)$ is stable and G satisfies $I(G) = I(f(\mathbf{w};\theta))$ 
  \parencite[p.31]{spirtes2000causation}
  \label{faithfulness}
\end{definition}

Other than the optional assumption of linearity and Gaussian errors that are made here for simplicity, and the assumed lack of unobserved confounders, \textit{stability} is the only assumption necessary for the identification of a DAG that represents a true DGP. This is the untestable component of the perhaps more commonly referenced concept of \textit{faithfulness}\footnote{Note that this definition of faithfulness includes an equivalence relationship and therefore encompasses what is sometimes referred to separately as the \textit{Causal Markov Condition} which states that $I(G) \subseteq I(f(\mathbf{w};\theta))$ \parencite{spirtes2016causal}}. Faithfulness as defined here also consists of the (testable) assumption that the graph $G$ captures all the conditional independence relationships in the DGP $f$\footnote{What it means for variables to be (conditionally) independent in a graph will be covered very shortly}. Stability is the assumption that these relationships are invariant to small perturbations in the parameters of the true DGP. Intuitively, if we wish to use conditional independence relationships to identify a model then we must assume that the observed conditional independence relationships do not belie the underlying relationships between variables. This assumption is violated only if some causal effects exactly cancel out, resulting in no observed correlation between casually connected variables. \citeauthor{pearl2009causality} (\citeyear{pearl2009causality}) provides the following example. Consider the following model: $z = \beta_{zx} x + \epsilon_x$, $ y = \beta_{yx} x + \beta_{yz} z + \epsilon_y$. If we impose the parameter restriction $\beta_{yx} = -\beta_{yz}\beta_{zx}$ then $x$ and $y$ are independent. However, this independence relationship is not robust to perturbations of the model parameters and is therefore not stable in the relevant sense. In this case the ground truth graph cannot be learned from observational data, as the causal dependence that exists between $x$ and $y$ is not captured by the conditional independence relationships in the data generated by this model. Under the assumption that $f$ is \textit{stable} we can use conditional independence tests in the following way to evaluate whether a DAG $G$ is consistent with (or more precisely \textit{faithful} to) $f$.

\begin{definition}{D-Separation}
  A path $P$ starting at node $x$ and ending at node $y$ in a DAG $G$ is said to be \textbf{d-separated} or \textbf{blocked} by a set of variables $\mathbf{z}$ if and only if the following two conditions hold: \\
  1. If $P$ contains a chain $x \rightarrow m \rightarrow y$ or fork $x \leftarrow m \rightarrow y$ then $m \in \mathbf{z}$ \textbf{and} \\
  2. If $P$ contains a collider $x \rightarrow m \leftarrow y$ then $\{m \cup des(m)\} \cap \mathbf{z} = \emptyset$ \\
  A set of variables $\mathbf{z}$ is said to \textit{d-separate} $x$ and $y$ if $\mathbf{z}$ blocks every path between $x$ and $y$.
  \parencite[p.16]{pearl2009causality}
  \label{dseparation}
\end{definition}

D-separation is sometimes also referred to in terms of \textit{backdoor} and \textit{frontdoor} paths \parencite{pearl2009causality}. A backdoor path is a path that links two nodes going back against the direction of an arrow from at least one node. Conversely, a frontdoor path is a path that travels down in the direction of the arrow from every node. Then a set of variables $\mathbf{z}$ can equivalently be said to D-separate nodes $x$ and $y$ if and only if $\mathbf{z}$ blocks all the backdoor paths from $x$ to $y$ (1.) and none of the frontdoor paths (2.).
 
\theoremstyle{theorem}
\begin{theorem}{D-Separation and Conditional Independence}
  If $x$ and $y$ are d-separated by $\mathbf{z}$ in DAG $G$, and $G$ is faithful to the true DGP $f(\mathbf{w};\theta)$ of $x$ and $y$, then $x$ and $y$ are independent conditional on $\mathbf{z}$. 
  \parencite[p.18]{pearl2009causality}
  \label{dseptheorem}
\end{theorem}

\theoremstyle{corollary}
\begin{corollary}{Test of Faithfulness}
  If $x$ and $y$ are d-separated in $G$ by $\mathbf{z}$ but $x$ and $y$ are not independent conditional on $\mathbf{z}$ in the true DGP $f(\mathbf{w};\theta)$, and $f(\mathbf{w};\theta)$ is stable then $G$ is not faithful to $f(\mathbf{w};\theta)$.
  \label{faithful_corollary}
\end{corollary}

The corollary is simply the negation of Theorem \ref{dseptheorem}, and it shows how the faithfulness of some DAG $G$ is falsifiable as long as $f$ is known to be \textit{stable}. This result is essential for defining the constraint-based tests in Section \ref{constrainttests}. In particular, it implies the following result that we will leverage:

\theoremstyle{corollary}
\begin{definition}{Parental Markov Condition}
  Given some DAG $G$, a node $x$ in $G$ is d-separated from and therefore independent of all its non-decedents by its parents. 
  \parencite[p.16, p.19]{pearl2009causality}
  \label{markovcompatibility}
\end{definition}

\theoremstyle{corollary}
\begin{corollary}
  If $G$ is faithful to the DGP $f$ over a set of observable variables $\mathbf{w}$ then $f$ admits the following factorisation:
  \begin{equation}
    f(\mathbf{w};\theta) = \prod_{i=1}^{k} f(w_i | pa_G(w_i);\theta)
  \end{equation} \parencite[p.16]{pearl2009causality}
\end{corollary}

\subsubsection{Estimation} \label{dag_estimation}

\begin{figure}
  \centering
  \includegraphics[width=0.75\textwidth]{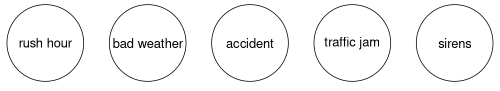}
  \caption{A DAG before structure learning}
  \label{dag2}
\end{figure}

There are two fundamental problems to solve when estimating a DAG. The first is known as \textit{parameter learning}, and the other \textit{structure learning} \parencite{ermon_2017}. Given a DAG as in Figure \ref{dag2}, the first task is simply to estimate the parameters of the network, such as the parameter matrices $\mathbf{A}$, $\mathbf{B}$, $\mathbf{C}$, $\mathbf{D}$, and $\mathbf{E}$ in equations (\ref{ss_solution:x}) - (\ref{ss_solution:z}) in Section \ref{dsge}. This is usually done via maximum likelihood or perhaps with Bayesian techniques.

The second and more onerous task, as demonstrated by Figure \ref{dag2} is that if we just start with some observational data and no model it is not obvious which edges between nodes need to be estimated in the first place. One way to do this is for the researcher to specify explicitly which edges should be present in the graph, and simply fit the parameters of that graph. As discussed in Section \ref{methodology}, this is straightforward to do for DSGE models, assuming the true state variables are known. However, doing so in this context would achieve little. This is equivalent to specifying a system of linear equations (VAR with some parameters restricted to zero) to be estimated, probably based on some economic model that was developed by other means. While this is then automatically encapsulated in a convenient, easily interpreted representation of the underlying assumptions, this approach does not offer anything particularly novel. 

Instead, a more promising approach is to algorithmically learn the structure of the graph, that is to learn a causal model, directly from observed data. One "brute force" method to solve this problem is to compute the posterior likelihood of every possible network, however, this number is super-exponential in the number of variables and, therefore it becomes very computationally expensive, very quickly \parencite{chickering1996learning}. As a response to this, many heuristic approximation techniques have been developed. These can be broadly grouped into two categories: constraint-based and score-based structure learning algorithms \parencite{spirtes1991algorithm} \parencite{verma1991equivalence}, which I will now briefly discuss in that order. 

\begin{figure}
  \centering
  \begin{subfigure}{0.3\textwidth}
    \centering
    \includegraphics[width=\linewidth]{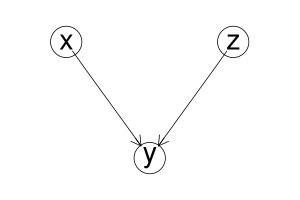} 
    \small
    \begin{equation*}
      x = \epsilon_{x}
    \end{equation*}
    \begin{equation*}
      y = \beta_{yx} x + \beta_{yz} z + \epsilon_{y}
    \end{equation*}
    \begin{equation*}
      z = \epsilon_{z}
    \end{equation*}
    \caption{Collider}
    \label{collider}
  \end{subfigure}
  \begin{subfigure}{0.3\textwidth}
    \includegraphics[width=\linewidth]{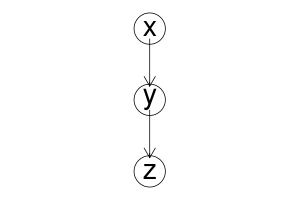}
    \small
    \begin{equation*}
      x = \epsilon_{x}
    \end{equation*}
    \begin{equation*}
      y = \beta_{yx} x + \epsilon_{y}
    \end{equation*}
    \begin{equation*}
      z = \beta_{zy} y + \epsilon_{z}
    \end{equation*}
    \caption{Chain}
    \label{chain}
  \end{subfigure}
  \begin{subfigure}{0.3\textwidth}
    \includegraphics[width=\linewidth]{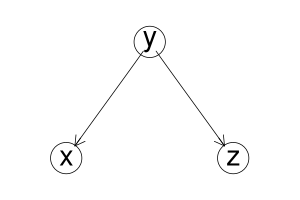}
    \small
    \begin{equation*}
      x = \alpha_x + \beta_{xy} y + \epsilon_{x}
    \end{equation*}
    \begin{equation*}
      y = \epsilon_{y}
    \end{equation*}
    \begin{equation*}
      z = \beta_{zy} y + \epsilon_{z}
    \end{equation*}
    \caption{Fork}
    \label{fork}
  \end{subfigure}

  \caption{The three possible v-structures of a 3 node DAG. Error terms $\epsilon$ are all i.i.d. Gaussian shocks.}
  \label{dag5}
\end{figure}

Constraint-based algorithms rely on the fact that changing the direction of an arc changes the conditional independence relationships implied by the graph, the presence of which can be tested for in the data. To see how the DAG assumptions can be sufficient to learn a causal model in this way, consider the example in Figure \ref{dag5}. Suppose we have a graph with three nodes, such that no one node is completely independent of the other two (as this would make the graph trivial, and we could rule out this case with an (unconditional) independence test). Furthermore, the graph cannot have all three possible arcs because it would either contain a cycle, or the third arc would imply a relationship which is redundant given the other two. Then the graph must have exactly two arcs. Given this, there are exactly three possible permutations of the network, which are the three shown in figure \ref{dag5}. These are known as the three canonical \textit{v-structures} \parencite{pearl2014probabilistic}. These structures are partially identifiable from observational data because they imply different testable hypotheses about conditional independence. While the chain and fork imply that x and z are unconditionally dependent and only independent conditional on y, the collider implies exactly the opposite; that x and z are unconditionally independent and dependent conditional on y. Given some observed data we can easily test for the presence of conditional and unconditional independence under the assumption of joint-normality using a t-test or F-test on (partial) correlations. The results of these tests can be used to rule out certain network structures which would be inconsistent with the observed data. Since this only separates on case from the other two, for every set of three variables the network is only partially identifiable, however, full identification can (but will not always) be achieved when more variables are observed. This is done by comparing overlapping triplets of variables and progressively reducing the set of network structures that are consistent both with the DAG assumptions and with the observed conditional independence relationships. There are many algorithms that have been implemented using this general approach, the most popular of which is the PC algorithm first developed by \citeauthor{spirtes1991algorithm} (\citeyear{spirtes1991algorithm}). This algorithm has been shown to consistently estimate (as $n \rightarrow \infty$) the structure of the ground truth DAG of observed data under the assumptions of linear and Gaussian conditional probability functions, stability, lack of unobserved confounders, and structural complexity that does not grow too quickly relative to $n$ \parencite{kalisch2007estimating}. 

Score-based\footnote{Here I use the meaning of "score" that is typical in the machine-learning literature --- some function to be maximised in order to improve model fit. This should not be confused with the common definition of "score" in the econometrics and statistics literatures, which is the gradient of the likelihood function.} methods assign some score to every network based on its predictive accuracy (usually related to the likelihood of the model) and then use (stochastic) gradient-descent\footnote{In this case to be precise it is really a gradient-\textit{ascent}.} to identify the optimal network structure. There are a number of functions and hill climbing algorithms that can be used to achieve this. In the case of continuous data the log-likelihood of the model or some penalised variant is usually used as the score function. A consistency result for the GES score-based algorithm is given in \citeauthor{chickering2002optimal} (\citeyear{chickering2002optimal}). The assumptions are slightly stronger than that of the PC algorithm --- the number of variables must be fixed rather than growing slowly relative to n.

The major benefit of the constraint-based method is that it directly utilises conditional independence as a primitive, which is the concept of causality that DAGs seek to identify. This is in contrast to score-based methods, which effectively maximise the predictive accuracy of the model, and there is seemingly no guarantee that the best predictive model is the most likely causal explanation. In other words, despite the presence of large sample consistency results for both types of algorithms, it seems reasonable to believe that bias due to finite samples or slight deviations from stated assumptions is likely to be more prominent for score-based methods. The major benefit of score-based methods on the other hand is that they will always converge to a single fully directed graph as a solution whereas constraint-based methods, because V-structures are only partially identifiable, may not be able to identify a unique solution. Instead, when the graph is only partially identifiable, the algorithm will return an undirected graph (CPDAG) \parencite{spirtes1991algorithm}. The undirected arcs in a CPDAG could face either direction and the graph would still be consistent with both the DAG assumptions and the observed conditional independences. By permuting the two possible directions of each undirected arc we arrive at a set of DAGs that are said to be \textit{observationally equivalent} or \textit{Markov equivalent} \parencite{colombo2014order}. This is problematic because it is difficult or impossible to fit parameters to and thereby derive counterfactual implications from graphs that are not fully directed.  

Fortunately, these two methods can be combined into so-called \textit{hybrid} structure learning methods which use the strengths of both methods to counter the weaknesses of the other \parencite{scutari2014multiple} \parencite{friedman2013learning}. In this method the algorithm maximises a score function, but the number of parents that each node can have is restricted. The main benefit of this is a large gain computation efficiency because the search space is dramatically reduced, and theoretically it has the benefits of both constraint-based and score-based learning. However, while the resulting graph is always directed, it does not always correctly reflect the observed v-structures because it trades off flexibly between constraint satisfaction and score maximisation (instead of giving lexicographic priority to constraint satisfaction, which is the approach that my algorithm will take). \citeauthor{nandy2018high} (\citeyear{nandy2018high}) give an asymptotic consistency result for a particular hybrid learning algorithm called ARGES.

\subsection{DSGE Models} \label{dsge}

Suppose a DSGE model is defined over a set of $k$ variables in a vector $\mathbf{w}_t$, for which one observation is available per time period, for example, quarterly or yearly data. The log-linear approximation to a stationary DSGE model solution can be written as a state-space model \parencite{king1988production} that partitions $\mathbf{w}_t$ into three mutually exclusive vectors $\mathbf{x}_t$, $\mathbf{y}_t$, and $\mathbf{z}_t$. This state-space model is defined by equations (\ref{ss_solution:x}) - (\ref{ss_solution:z}):

\begin{align}
  \mathbf{y}_t &= \mathbf{A} \mathbf{x}_{t-1} + \mathbf{B} \mathbf{z}_{t} \label{ss_solution:x}\\
  \mathbf{x}_t &= \mathbf{C} \mathbf{x}_{t-1} + \mathbf{D} \mathbf{z}_{t} \label{ss_solution:y}\\
  \mathbf{z}_t &= \mathbf{E} \mathbf{z}_{t-1} + \mathbf{\epsilon}_{t} \label{ss_solution:z}
\end{align}

Where $\mathbf{x}_t$ is a vector of endogenous state variables, $\mathbf{y}_t$ is a vector of control variables, $\mathbf{z}_t$ is a vector of exogenous state variables, $\mathbf{A}$, $\mathbf{B}$, $\mathbf{C}$, $\mathbf{D}$, and $\mathbf{E}$ are coefficient matrices, and $\mathbf{\epsilon}_t$ is a vector of shocks. All variables are mean-zero. The shocks in $\mathbf{\epsilon}_t$ can be interpreted as structural shocks as they satisfy the assumptions $\mathbf{\epsilon}_{t} \sim N(0,\Sigma)$ and $\Sigma$ diagonal $\implies Cov[\epsilon_{i,t},\epsilon_{j,t}] = 0 \iff \epsilon_{i,t} \indep \epsilon_{j,t} $ for $i \not = j$. These shocks are assumed to not be observed, both because this is likely true in realistic applications (absent some very clever econometric tricks) and because observing the shocks is simply not necessary for the type of inference proposed in this paper.

Furthermore, assume that $\mathbf{E}$ is diagonal ($e_{ij} = 0$ if $i \not = j$) such that the process of each exogenous state depends only on its own past and $|e_{ii}| < 1$ such that the model is stationary. Note that this structure implies that the exogenous states possess the Markov property, that is, $\mathbf{z}_t$ depends only on $\mathbf{z}_{t-1}$ and not any further lags. As a result, the entire model has the Markov property. However, the framework and algorithm proposed here could in principle be generalised to allow for longer lags, if for example, it is believed that some effects may take multiple periods to play out.

In this setup, all variables can be categorised as either state variables or control variables \parencite{fernandez2016solution}. Defined as broadly as possible, state variables are the variables whose past is relevant for determining the current value of modelled variables, and control variables are everything else; their past is irrelevant to the current values of the model. State variables can be further categorised as either endogenous states (the capital stock in the economy is a typical example) and exogenous states (the state of technology or productivity is a typical example) \parencite{ravenna2007vector}. As the name suggests, endogenous states are determined simultaneously (endogenously) with contemporaneous controls in the model, however, their past is relevant to the determination of the current values of the model. Exogenous states, on the other hand, are exogenous in the strongest possible sense. In this setup they are \textit{strictly exogenous} relative to and not Granger caused by any other variable in the model, including the other exogenous states.

\section{Methodology} \label{methodology}

\begin{figure}
  \centering
  \begin{tikzpicture}[scale=1.5]
    \tikzset{
      vertex/.style={circle,draw, minimum size=2em},
      edge/.style={->,> = latex'}
    }
    \node[vertex] (zt) at (1 ,0) {$\mathbf{z}_{t\quad}$};
    \node[vertex] (zt1) at (1 , 1) {$\mathbf{z}_{t-1}$};
    \node[vertex] (zt2) at (1 , 2) {$\mathbf{z}_{t-2}$};
    \node[vertex] (xt) at (0 ,-1) {$\mathbf{x}_{t\quad}$};
    \node[vertex] (xt1) at (0 , 0) {$\mathbf{x}_{t-1}$};
    \node[vertex] (xt2) at (0 , 1) {$\mathbf{x}_{t-2}$};
    \node[vertex] (yt) at (-1,-1) {$\mathbf{y}_{t\quad}$};
    \node[vertex] (yt1) at (-1, 0) {$\mathbf{y}_{t-1}$};
    \node[vertex] (yt2) at (-1, 1) {$\mathbf{y}_{t-2}$};
    \node[vertex] (et) at (2 , 0) {$\mathbf{\epsilon}_{t\quad}$};
    \node[vertex] (et1) at (2 , 1) {$\mathbf{\epsilon}_{t-1}$};
    \node[vertex] (et2) at (2 , 2) {$\mathbf{\epsilon}_{t-2}$};

    \draw[edge] (et) -- (zt);
    \draw[edge] (et1) -- (zt1);
    \draw[edge] (et2) -- (zt2);
    \draw[edge] (zt2) -- (zt1);
    \draw[edge] (zt1) -- (zt);
    \draw[edge] (xt2) -- (xt1);
    \draw[edge] (xt1) -- (xt);
    \draw[edge] (zt2) -- (xt2);
    \draw[edge] (zt1) -- (xt1);
    \draw[edge] (zt) -- (xt);
    \draw[edge] (zt2) -- (yt2);
    \draw[edge] (zt1) -- (yt1);
    \draw[edge] (zt) -- (yt);
    \draw[edge] (xt2) -- (yt1);
    \draw[edge] (xt1) -- (yt);
  \end{tikzpicture}
  \caption{DSGE solution expressed as a DAG}
  \label{dsge_dag}
\end{figure}
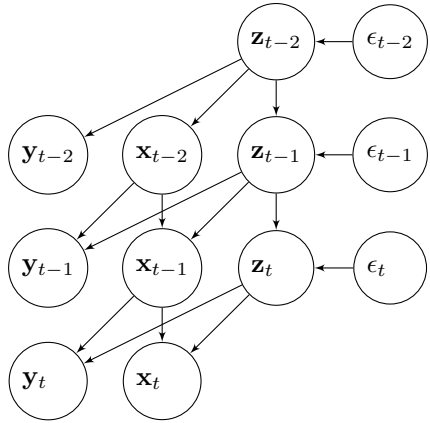

Given equations (\ref{ss_solution:x}) - (\ref{ss_solution:z}) it is straightforward to characterize the general solution to a DSGE model as a DAG. This is illustrated by Figure \ref{dsge_dag}. This expresses in graphical format all the relationships that exist in those equations, taking into consideration that the edges in the DAG are assumed to imply linear relationships. In particular, it captures all the conditional independence relationships in the DSGE model, therefore, if the underlying distribution is stable, then this DAG is faithful. 

Already it would seem straightforward to input random samples generated from a DSGE model into the available structure learning algorithms in order to find the correct model, given that these algorithms have well established asymptotic convergence properties. Unfortunately, results obtained this way (provided in Sections \ref{rsmax2} and \ref{pc}) are less than convincing, as these algorithms seem to have a number of important limitations in this context. Constraint-based algorithms rely on conditional independence tests which themselves involve computing the correlation between residuals. In the context of simulated data these residuals may be very small or effectively zero when conditioning on the true parents of a variable. While correlation is undefined for constants, in practice when the calculation is forced the result tends to infinity, as it involves division by a number very close to zero. This is particularly problematic because the case where the true parents are conditioned on is exactly the case in which we wish to conclude that the remaining partial correlation is in fact zero. Furthermore, these results are only asymptotic, and it seems that finite-sample bias may be important in economic applications, where in practice sample sizes are small relative to the dimension of the problem. Particularly problematic is that structure learning algorithms consider all possible DAGs given observed variables as potential candidates, whereas in this context we assume that the solution takes on a particular form, as in equations (\ref{ss_solution:x}) - (\ref{ss_solution:z}).

As a result of these limitations, I found that a more effective approach in this context involved a bespoke algorithm that takes into account the relatively stringent assumptions that can be made about DSGE solutions. For the reasons outlined in Section \ref{dag_estimation} this will be a hybrid algorithm. Therefore, before introducing the algorithm I will define relevant constraint and score tests in turn. But first, I will discuss the validity of the \textit{stability} assumption (Definition \ref{stability}), which is essential for any DAG estimation procedure to be justified, in the context of DSGE models. Note that here we will be discussing stability which is the fundamental untestable assumption under the definitions given in this paper, however, often in the DAG literature it is instead the faithfulness assumption which is discussed. This is because when discussing the appropriateness of a specific DAG faithfulness is the relevant concept. In order to be faithful, the DAG and DGP must satisfy both stability and have the same conditional independence relationships. However, at the moment we are more interested in whether any appropriate DAG exists, which depends only on stability. If stability holds we can then search for a DAG that also satisfies the testable part of faithfulness using particular conditional independence tests that will be introduced in Section \ref{constrainttests}.

\subsection{Validity of the Stability Assumption}

A sufficient condition for stability is that the DGP parameters are jointly continuous and vary freely over the parameter space \parencite{steel2006homogeneity} for different populations, or equivalently, that the matrix of DGP parameters is of full rank. This is because under this condition, specific combinations of parameters which result in the cancellation of causal effects as in the example in Section \ref{prelim} have Lebesgue measure 0. If we believe that the true DGP of the macroeconomy is a DSGE model, which itself is faithfully represented by a DAG, then this condition is unlikely to be met. DSGE models impose many cross-equation restrictions on parameters that effectively reduce the rank of the parameter matrix. Unfortunately this condition will not allow us to guarantee that DSGE models satisfy the stability assumption. Regardless, this condition is merely sufficient, not necessary, and so it does not rule out that DSGE models can be faithfully represented by DAGs. 

In another approach to failures of stability, \citeauthor{steel2006homogeneity} (\citeyear{steel2006homogeneity}) notes that such failures or near-failures (that is near-zero statistical dependence despite clear causal pathways) are likely to occur when parameters are both subject to \textit{selection} and \textit{homogeneity}. In this context, selection means that parameters are entirely determined by an economic agent. The suggestion is that if the path of a policy variable $z$ is specifically designed as a function of $x$ to counteract the causal effect of $x$ on some outcome $y$, then it is reasonable to believe that little or no correlation will be observed between $x$ and $y$ despite a clear causal pathway between them. If parameters are assumed to come from some distribution with different draws for each population, then \textit{homogeneity} is the statement that there is little exogenous variation in those parameter values, that is variation outside the variation caused by selection. This condition is perhaps more likely to be violated as there is considerable cross-country variation in macroeconomic conditions. However again, surviving the selection and homogeneity test does not guarantee that the stability assumption is verified.

Despite these concerns, I would argue that the stability assumption is plausible in most macroeconomic contexts. For simulations, whether or not the assumption is violated can be read straight off the model. For real data, it seems unlikely that any macroeconomic variable (even the policy rate) is determined in an entirely systematic or deterministic way. In reality, monetary authorities face a number of constraints that would prevent them from completely stabilising inflation including informational constraints, political influences, and the zero lower bound. Identification of policy rate shocks has been a topic of much scrutiny \parencite{ramey2016handbook}, and this line of research has provided a significant amount of evidence for the existence of such shocks, suggesting that even in the context of monetary policy where decision-making is highly rigorous, it is nonetheless neither strictly deterministic, nor determined in the same way everywhere. For other macroeconomic variables which are not determined by a centralised authority, it is even easier to believe there is significant room for free variation of parameters across populations.

\subsection{Constraint Tests} \label{constrainttests}

\subsubsection{Independence Relationships}

 Applying the parental Markov condition (Corollary \ref{markovcompatibility}) to the DAG in Figure \ref{dsge_dag} implies the following four independence relationships among the time $t$ and $t-1$ variables:

\begin{align}
  x_t \indep x^{\prime}_{t} \,||\,& [\mathbf{x}_{t-1},\mathbf{z}_t] \text{ for all } x_t \neq x^{\prime}_{t} \in [\mathbf{x}_t, \mathbf{y}_t] \label{constraint_test:1} \\
  x_{t-1} \indep z_{t} \,||\,& \mathbf{z}_{t-1} \text{ for all } x_{t-1} \in \mathbf{x}_{t-1} \text{ and } z_{t} \in \mathbf{z}_t \label{constraint_test:3} \\
  x_t \indep z_{t-1} \,||\,& [\mathbf{x}_{t-1}, \mathbf{z}_t] \text{ for all } x_t \in [\mathbf{x}_t, \mathbf{y}_t] \text{ and } z_{t-1} \in \mathbf{z}_{t-1} \label{constraint_test:2} \\
  z_t \indep z^{\prime}_{t} || \mathbf{z}_{t-1} & \text{ for all } z_t \not = z^{\prime}_{t} \in \mathbf{z}_t \label{constraint_test:4}
\end{align}

The first condition (\ref{constraint_test:1}) is the statement that the model's time $t$ endogenous variables are explained entirely by and are therefore unconfounded conditional on $\mathbf{x}_{t-1}$ and $\mathbf{z}_t$ (I sometimes hereafter refer to these as the time $t$ states). In DAG parlance, a time $t$ endogenous variable is \textit{d-separated} from and therefore independent of any other time $t$ endogenous variable by the time $t$ states. Condition (\ref{constraint_test:3}) states that the time $t$ lagged endogenous states are independent of every exogenous state conditional only on the lagged exogenous states. This follows from the exogeneity of $\mathbf{z}$ which implies that the only parent of $z_t$ other than the shock is $z_{t-1}$. Condition (\ref{constraint_test:2}) holds because the time $t$ states d-separate the time $t$ endogenous variables from the lagged exogenous states. If we were to consider further lags, this conditional independence would apply not only to $z_{t-1}$, but also to all $t-2$ and earlier variables because of the Markov condition. Finally, Condition (\ref{constraint_test:4}) holds that all exogenous states are mutually independent conditional on past exogenous shocks. This is a stronger condition than the other three, and depends crucially on the assumptions that $\mathbf{E}$ and $\Sigma$ are diagonal.

We consider (conditional) only independence relationships because it is usually the case in macroeconomic timeseries that all observables have non-zero pairwise correlation. Therefore, the \textit{lack} of a relationship in the form of conditional independence is more salient and therefore is more useful for identification than the presence of one. We can apply Corollary \ref{faithful_corollary} to these constraint tests in order to already provide a powerful selection criteria for empirical DSGE models, which I will refer to as \textit{validity}:

\theoremstyle{theorem}
\begin{theorem}
  Suppose that a log-linearised DSGE model $M$ generates a DGP $f(\mathbf{w}_t;\theta)$ over a set of observed variables $\mathbf{w}_t$, that partitions $\mathbf{w}_t$ into three mutually exclusive vectors $\mathbf{x}_t$, $\mathbf{y}_t$, and $\mathbf{z}_t$ representing the endogenous states, controls, and exogenous states of $M$ respectively. Further suppose that there is some DAG $G$ which is \textit{faithful} to $f(\mathbf{w}_t;\theta)$. Then $G$ is the only faithful DAG which satisfies conditions (\ref{constraint_test:1}), (\ref{constraint_test:3}), and the minimum state variable criterion (MSV) \parencite{mccallum1999role}. This $G$ is said to be \text{valid}.
  \label{validity}
\end{theorem}

Proof of this theorem is provided in Appendix \ref{validity_proof}. The introduction to Section \ref{methodology} showed that under the assumption of stability a log-linear DSGE solution can be represented by at least one faithful DAG, therefore by applying the previous theorem, we arrive at the following corollary:

\theoremstyle{corollary}
\begin{corollary}
  If an underlying DSGE model which can be expressed as in equations (\ref{ss_solution:x}) - (\ref{ss_solution:z}) generates a \textit{stable} distribution $f(\mathbf{w}_t;\theta)$, then there is exactly one valid DAG which is faithful to $f(\mathbf{w}_t;\theta)$.
  \label{unique_dag}
\end{corollary}

Note that while the proof of Theorem (\ref{validity}) makes use of only constraints (\ref{constraint_test:1}) and (\ref{constraint_test:3}), (\ref{constraint_test:2}) and (\ref{constraint_test:4}) are still applicable (necessary conditions) because they are implied by the DAG, but they are not in the minimal set of sufficient conditions for a unique solution. To be more general we could drop these assumptions as long as the shocks only directly affect the exogenous states, and the other constraints would still hold and be valid tests of the model. However, these constraints (and the associated assumptions) can nonetheless be included because they are satisfied by a wide range of DSGE models including all of those considered in the empirical portion of this paper, and more importantly testing a larger number of conditions will, all else equal, give more \textit{power} to reject incorrect models, which will prove critical in dealing with issues arising from small available sample sizes.

The Minimum State Variable (MSV) criterion simply states that the chosen model should have the fewest number of state variables among those which satisfy the conditional independence criteria. This criterion is necessary for the proof, however, it is a natural and intuitive requirement to impose. Adding state variables increases the size of the conditioning set, and therefore weakly increases the plausibility of any conditional independence relationship we may test\footnote{To see this, note that $\mathbf{y}_{t-1}$ has no children, so $y_{t-1} \in \mathbf{y}_{t-1}$ cannot result in a blocked frontdoor path if it is moved into $\mathbf{x}_{t-1}$, while no new conditional independence restrictions that must be satisfied are implied.} As a result, if some model with $m$ states is valid, then another model with the same states save for one control that is changed to a state variable will also trivially be valid. Yet that model would also be less parsimonious and is therefore less desirable. This can be seen as the application of \textit{Occam's Razor} to state-space models, wherein state variables have more complex dynamics than controls. Consider equations (\ref{ss_solution:x}) - (\ref{ss_solution:z}). Exogenous states are involved in all three equations, endogenous states two, and controls only one. Another way to see this is in figure \ref{dsge_dag}. Among time $t$ and $t-1$ variables, adding an exogenous state results in the addition of edges in four places and thus eight parameters (one slope parameter and one variance parameter), an endogenous state in three places, and a control in only two. Therefore, according to this principle, models with fewer states, especially exogenous states are preferable, all else equal. 

It is worth stopping now to consider what Corollary \ref{unique_dag}, and the results leading up to it, collectively imply. Consider some distribution $f(\mathbf{w}_t,\theta)$, and assume that all the conditional independence relationships implied by it ($I(f(\mathbf{w}_t,\theta))$) are known. Furthermore, assume that $f(\mathbf{w}_t,\theta)$ is \textit{stable}, and was generated by some sort of log-linear DSGE model as in equations (\ref{ss_solution:x}) - (\ref{ss_solution:z}), whose exact specification is unknown. Then there is exactly one DAG, and hence state-space model that is \textit{valid}. It is then possible to identify this state-space model by testing all possible state-space models against the validity criteria. While this does not uniquely pin down a single set of microfoundations, it does rule out any structural model that does not generate the correct state-space in its reduced form. This shows the degree of identification that we are at least theoretically able to obtain using this approach, however, it will be necessary to consider many practical issues in order to operationalise this, including how to test for conditional independence, which is the issue that will be considered next.

\subsubsection{Testing Procedure} \label{testing}

The proof in Theorem \ref{validity} assumed that we could observe the conditional independence relationships in the true distribution of observables. Of course, this is not the case in practice, therefore, this section will discuss the implementation of an empirically viable strategy for testing conditions (\ref{constraint_test:1}) - (\ref{constraint_test:4}). In the present application, I make the assumption that observed variables are normally distributed, such that testing for conditional independence is equivalent to testing for non-correlation among partial residuals. This assumption is in general not required as it is possible to test for conditional independence non-parametrically (see \citeauthor{strobl2019approximate} (\citeyear{strobl2019approximate}) for a review of recent contributions in this vein), however, this assumption is made here because Gaussian assumptions are common in DSGE models and economic applications more generally, and the resulting simplifications will allow for more clear exposition of the main contributions of this paper. 

Partial linear correlations can be estimated by regressing the set of target variables of interest $\mathbf{x}$ on the set of conditioning variables $\mathbf{z}$ and then estimating the correlations between the resulting estimated residuals $\hat{\mathbf{u}_x}$. Therefore, one way to implement tests for conditions (\ref{constraint_test:1}) - (\ref{constraint_test:4}) would be to perform a t-test on the estimated partial linear correlation implied by each of these conditions for every model, and then reject the model if any of these t-tests reject the null hypothesis at the specified significance level (after applying a \citeauthor{bonferroni1936teoria} (\citeyear{bonferroni1936teoria}) correction). Hereafter this is referred to as the \textit{multiple testing approach}. As shown in Section \ref{results}, this approach does seem to perform well on simulated data, with higher power and lower size than the second approach which I will soon introduce. However, it has a number of significant drawbacks. 

Firstly, the \citeauthor{bonferroni1936teoria} (\citeyear{bonferroni1936teoria}) correction assumes independence of each of the tests, which is highly implausible in this case. Indeed, this explains why the empirical size of these tests is less than the specified significance level. Since the degree of correlation between tests may take on any form it is difficult or impossible to pin down important statistical properties (such as the size or power) of this procedure. Furthermore, there is the issue (which was also noted as a drawback of alternative approaches) that computation of partial correlations can be unstable if residuals are very close to or equal to zero. Indeed, in principle the residuals produced by the correct model should be exactly zero, which is a constant, and therefore pairwise correlation undefined. In practice, when simulated data is used, residuals for the ground truth model are very close to (but not equal to) zero. In this case pairwise correlation can be computed, however, it is not particularly meaningful since it only reflects floating point imprecision or rounding error in the simulation. Since this computation involves dividing two near-zero values it tends to produce an estimated correlation close to 1. This is problematic because this is exactly when we do not want to reject the null hypothesis of conditional independence. As a workaround for this the algorithm will detect small residuals below some tolerance threshold and pass the model through the test (do not reject the hypothesis of independence) if they are observed. This is a highly idiosyncratic correction that is an undesirable feature of this approach. Finally, the number of tests conducted can grow very large if there is a large number of observables resulting in implausibly large critical values (due to the Bonferroni correction), and exponentially growing computational complexity.

For these reasons, I also propose the implementation of a different test provided by \citeauthor{srivastava2005some} (\citeyear{srivastava2005some}). This test is for the null hypothesis that a covariance matrix is diagonal. In order to use this, we will combine and slightly rearrange conditions (\ref{constraint_test:1}) - (\ref{constraint_test:4}) such that they have the same conditioning set, and imply a relationship of \textit{complete partial independence} between tested variables. To do this, roll conditions (\ref{constraint_test:1}) and (\ref{constraint_test:2}) back one period\footnote{Equivalently, one could roll forwards the other two conditions, but this would require data on a lead rather than two lags.}, and add $\mathbf{x}_{t-2}$ to the conditioning sets in conditions (\ref{constraint_test:3}) and (\ref{constraint_test:4}). This latter change is justified because in both cases we have already blocked every backdoor path between the variables of interest and $\mathbf{x}_{t-2}$ is not part of any frontdoor path between them, and therefore d-separation is maintained. In other words, if the exogenous states are mutually independent, conditioning on endogenous states contains no new relevant information, so it is harmless to add these to the conditioning set. The modified conditions are shown in (\ref{mod_constraint_test:1}) - (\ref{mod_constraint_test:4}).

\begin{align}
  x_{t-1} \indep x^{\prime}_{t-1} \,||\,& [\mathbf{x}_{t-2},\mathbf{z}_{t-1}] \text{ for all } x_{t-1} \neq x^{\prime}_{t-1} \in [\mathbf{x}_{t-1}, \mathbf{y}_{t-1}] \label{mod_constraint_test:1} \\
  x_{t-2} \indep z_{t-1} \,||\,& [\mathbf{x}_{t-2},\mathbf{z}_{t-1}] \text{ for all } x_{t-1} \in \mathbf{x}_{t-1} \text{ and } z_{t-1} \in \mathbf{z}_{t-1} \label{mod_constraint_test:3} \\
  x_{t-1} \indep z_{t-2} \,||\,& [\mathbf{x}_{t-2},\mathbf{z}_{t-1}] \text{ for all } x_{t-1} \in [\mathbf{x}_{t-1}, \mathbf{y}_{t-1}] \text{ and } z_{t-2} \in \mathbf{z}_{t-2} \label{mod_constraint_test:2} \\
  z_{t-1} \indep z^{\prime}_{t-1} || & [\mathbf{x}_{t-2},\mathbf{z}_{t-1}] \text{ for all } z_{t-1} \not = z^{\prime}_{t-1} \in \mathbf{z}_{t-1} \label{mod_constraint_test:4}
\end{align}

We now have that each of the conditional independence relationships (\ref{mod_constraint_test:1}) - (\ref{mod_constraint_test:4}) relies on the same conditioning set. Furthermore, when combined these conditions imply that all the variables in the vector $[\mathbf{y}_{t-1}, \mathbf{x}_{t-1}, \mathbf{z}_{t}, \mathbf{z}_{t-2}]$ are completely independent, conditional on $[\mathbf{x}_{t-2}, \mathbf{z}_{t-1}]$. $\mathbf{z}_{t-2}$ can be optionally excluded from this vector\footnote{In the application this will be excluded because the independence of $\mathbf{z}_{t-2}$ here depends on the strict exogeneity property of the exogenous states rather than the mutual independence of their AR processes (since we do not condition on $\mathbf{z}_{t-3}$). This may lead to some false rejections of the true model because of the possibility of a spurious regression when the exogenous states are close to unit roots.} as it is associated with test (\ref{constraint_test:2}), which is not required for a unique \textit{valid} model as in Theorem \ref{validity}. On the other hand we will have to impose (\ref{constraint_test:4}), which is also not required for \textit{validity}, in order to implement this test. This condition is now necessary because without it the partial covariance matrix would have some unrestricted elements and not be strictly diagonal under the null hypothesis, which would therefore be a substantially more difficult hypothesis to test. To test whether a model is valid using this approach, we first regress the vector $[\mathbf{y}_{t-1}, \mathbf{x}_{t-1}, \mathbf{z}_{t}]$ on $[\mathbf{x}_{t-2}, \mathbf{z}_{t-1}]$, collect estimated residuals, estimate the covariance matrix $S$ of the $T \times k$ matrix of residuals and, perform a z-test at some specified significance level $\alpha$ on the test statistic $\hat{T}_3$ from \citeauthor{srivastava2005some} (\citeyear{srivastava2005some}), which is asymptotically normally distributed. This test statistic is defined by equations (\ref{t3}) - (\ref{a40}):

\begin{align}
  \hat{T}_3 = \left(\frac{n}2{}\right)\frac{(\hat{\gamma}_3-1)}{\left(1-\left(\frac{1}{p}\right)\left(\frac{\hat{a}_{40}}{\hat{a}^2_{20}}\right)\right)^\frac{1}{2}} \label{t3} \\
  \hat{\gamma}_3 = \frac{n}{n-1}\frac{tr(S^2) - \frac{1}{n}(tr(S))^2}{\sum_{i=1}^ms^2_{ii}} \\
  \hat{a}_{20} = \frac{n}{p(n+2)}\sum_{i=1}^ms^2_{ii} \\
  \hat{a}_{40} = \frac{1}{p}\sum_{i=1}^ms^4_{ii} \label{a40}
\end{align}

Note that $s_{ij}$ is the $(i,j)$ element of $S$. Also note that the denominator in $\hat{T}_3$, $\left(1-\left(\frac{1}{p}\right)\left(\frac{\hat{a}_{40}}{\hat{a}^2_{20}}\right)\right)$ can be negative, and thus, the test statistic undefined. In order to alleviate this I take the same approach as in \citeauthor{wang2013necessary} (\citeyear{wang2013necessary}) and replace this term with $1-\sum_{i=1}^ms^4_{ii}/\left(\sum_{i=1}^ms^2_{ii}\right)^2$ when it is negative.

This strategy alleviates a number of the drawbacks of the first approach. Since this approach utilises estimated covariance rather than correlations it avoids unstable computation around the true model, where residuals are very close to zero. Therefore, we are able to test all models without making exceptions for special cases. Furthermore, it is much simpler to describe the properties of this test. Asymptotically, it will have exactly $\alpha$ type I error rate, without the need for any corrections. Estimates of the power of this test against numerous alternatives can be found in \citeauthor{wang2013necessary} (\citeyear{wang2013necessary}), and are also provided over a range of scenarios in Appendix \ref{testing_validation}. Finally, this approach results in exactly one test being performed regardless of the complexity of the model under consideration. While it is true that the test is somewhat more computationally intensive for larger covariance matrices, it has much better computational scaling properties than the multiple testing strategy. However, as we will see in Section \ref{results}, it unfortunately does not seem to be as accurate at identifying the ground truth model in simulations on more complex data sets as the multiple testing strategy. This seems to be due to a lack of power against alternatives.

\subsection{Score Tests} \label{scoretests}

Notwithstanding the uniqueness proof in Theorem \ref{validity}, in finite samples it is not uncommon to encounter cases where more than one model is \textit{valid}. The results in Section \ref{results} show that these models will generally be very similar to the ground truth, and represent a small minority of all considered models. At this point it could be left up to expert opinion to select the most sensible of the remaining models, however, since one of the most important benefits of this approach is agnosticism it is desirable to implement some heuristic way of selecting a single model with the algorithm. The approach that I will use to achieve this is score maximisation over valid models. Essentially, this will sort the models which are deemed to be valid by their likelihood in order to choose a unique winning model. In principle, one could evaluate models solely on their score, however, for the reasons outlined in Section \ref{dag_estimation}, my preferred approach is to use this only in a secondary role. For comparison simulation results for pure score-based estimation will be considered in Section \ref{score}.

The most basic score function for Gaussian Bayesian networks is the log-likelihood function. According to the parental Markov condition (Definition \ref{markovcompatibility}) if DAG $G$ is faithful to the DGP $f$, then $f$ admits factorisation of the joint probability distribution into the product of the distribution of each variable conditional on its parents:

\begin{equation}
  f(\mathbf{w};\theta) = \prod_{i=1}^{k} f(w_i | pa_G(w_i);\theta)
\end{equation}

Therefore, the log-likelihood can be calculated as:

\begin{equation}
  \mathcal{L}(\mathbf{w},\theta) = \sum_{i=1}^{k} ln(f(w_i | pa_G(w_i);\theta))
\end{equation}

Now consider the assumptions in the current context. $\mathbf{w}$ is partitioned into $\mathbf{z}$, $\mathbf{x}$, and, $\mathbf{y}$. We assume that the conditional probabilities are linear functions and follow a mean-zero normal distribution, so the only parameters are the slope parameters in matrices $\mathbf{A}$ - $\mathbf{E}$ and the variance-covariance matrix $\tilde{\Sigma}$. Furthermore, the model predicts time $t$ values \textit{given} time $t-1$ values, so we do not need to consider the distribution of lags which are constant with respect to the model. Therefore,

\begin{align}
  \mathcal{L}(\mathbf{w}; \mathbf{A}, \mathbf{B}, \mathbf{C}, \mathbf{D}, \mathbf{E}, \tilde{\Sigma}) =& \sum_{z_{i,t} \in \mathbf{z}_t} \bigg{(} \sum_{t=1}^{T} ln(\phi(z_{i,t} | z_{i,t-1} | \mathbf{E}, \tilde{\Sigma}_z)) \bigg{)} + \nonumber\\
  & \sum_{y_{i,t} \in [\mathbf{y}_t,\mathbf{x}_t]} \bigg{(} \sum_{t=1}^{T} ln(\phi(y_{i,t} | [\mathbf{x}_{t-1}, \mathbf{z}_{t}] | \mathbf{A}, \mathbf{B}, \mathbf{C}, \mathbf{D}, \tilde{\Sigma}_y)) \bigg{)} \\
  =& \sum_{i | y_{i,t} \in \mathbf{y}_t} \bigg{(} \sum_{t=1}^{T} ln(\phi(\mathbf{a}_i \mathbf{x}_{t-1} + \mathbf{b}_i \mathbf{z}_t | \mathbf{a}_i, \mathbf{b}_i, \sigma_i^2)) \bigg{)} + \nonumber\\
  & \sum_{i | x_{i,t} \in \mathbf{x}_t} \bigg{(} \sum_{t=1}^{T} ln(\phi(\mathbf{c}_i \mathbf{x}_{t-1} + \mathbf{d}_i \mathbf{z}_t | \mathbf{c}_i, \mathbf{d}_i, \sigma_i^2)) \bigg{)} + \nonumber\\
  & \sum_{i | z_{i,t} \in \mathbf{z}_t} \bigg{(} \sum_{t=1}^{T} ln(\phi(\mathbf{e}_i z_{i,t-1}  | \mathbf{e}_i, \sigma_i^2)) \bigg{)}
\end{align}

Where $\mathbf{x_i}$ is the $i_{th}$ row of $\mathbf{X}$, $\sigma_i^2$ are the $(i, i)$ diagonal elements of $\tilde{\Sigma}$, and $\phi$ is the probability density function of the normal distribution. Notice that we can calculate the variances separately in each linear projection because the parental Markov condition implies that each regression equation is independent. Therefore, we can substitute in for the maximum likelihood estimate of $\sigma_i^2$ for each regression and the functional form of $\phi$ to arrive at a substantially simpler expression for the log-likelihood function:

\begin{align}
  \mathcal{L(\mathbf{w})} =& -\frac{T}{2} \left( 
  k \left( 1 + ln(2\pi) \right) +
  \sum_{i=1}^{k} ln(\hat{\sigma}_{i}^2)
  \right) \\
  \hat{\sigma}_{i}^2 =& \frac{1}{T} \sum_{t=1}^{T} (w_{i,t} - \hat{w}_{i,t})^2
\end{align}

Where $\hat{w}_{i,t}$ are the predicted values of some $w_i$ in $\mathbf{w}$ implied by equations (\ref{ss_solution:x}) - (\ref{ss_solution:z}) using maximum likelihood estimates of the coefficient matrices. Note that the log-likelihood is inversely proportional to the mean squared error (MSE). This is consistent with the interpretation that maximising the score function is equivalent to finding the model with the best predictive performance in this context. Indeed, since the rest of the terms are constant, it suffices to minimise the MSE to maximise the log-likelihood in this setup.

Since maximising the log-likelihood does not penalise complexity, it often favours models with many more edges than exist in the ground truth. In other words, maximising log-likelihood over a space of candidate DAGs may lead to \textit{overfitting}. The most common response to this is to use a penalised score function such as the Akaike Information Criterion (AIC) \parencite{akaike1974new} and the Bayesian Information Criterion (BIC) \parencite{schwarz1978estimating}, which will regularise the estimated model by reducing the score by some increasing function of the number of parameters estimated. Indeed, in their proof for score-based GES algorithm consistency, \citeauthor{chickering2002optimal} (\citeyear{chickering2002optimal}) require that the score function used adequately penalise complexity.

When it comes to the preferred hybrid algorithm proposed here, given that we are already applying stringent conditional independence criteria, it may seem that this bias towards complexity is irrelevant. However, given the minimal number of states, it is still possible to reallocate between exogenous and endogenous states. In this context the bias towards complexity means that we are likely to choose more exogenous states than truly exist, since these involve the estimation of more parameters than endogenous states, and since they enter at time $t$ instead of time $t-1$ they likely contain more relevant information about time $t$ endogenous variables. In experimentation, I found that penalised score functions are very unlikely to overturn this bias towards exogenous states. So instead of using these, I will simply take lexicographic preference for models (among those which are valid) with more endogenous states first, and then only after this maximise the likelihood function. With this sorting all remaining models have the same complexity so penalised scores no longer serve any purpose. Another justification for this sorting is that in macroeconomics we generally believe that all observables are interrelated in some way, and therefore, the exogeneity assumptions implied by exogenous states are quite strong, and it is thus preferable to minimise them. 

\subsection{Algorithm} \label{algo}

Having defined a number of tests for an optimal and \textit{valid} model, we now turn our attention to developing an algorithm which will apply these tests in order to choose one from the set of all possible state-space models, which is outlined in Algorithm \ref{constraint_algo}.

The algorithm is very simple and is designed to reflect a few key model selection heuristics. As previously discussed, the algorithm assumes that constraint validity is more important than score maximisation. The scores of models that are not valid relative to the constraints are irrelevant because these models are thrown out. The justification for this is outlined in \ref{dag_estimation}. Essentially, unlike score functions, constraints directly rely on information about a relevant sense of causality.

The MSV criterion is imposed by the algorithm in the sense that it stops considering models with a greater number of states once some valid model is found. This is primarily because MSV is required for the uniqueness property of validity, however, the MSV criterion also allows for a potentially very large increase in the computational speed of the algorithm. Without it we must consider every possible combinations of states. Since the choice of states is multinomial with three categories, the complexity of this algorithm is $\mathcal{O}(3^k)$. However, if the ground truth has only $m < k$ states then we can skip $\sum_{r=m}^{k} 2^r {\binom{k}{r}}$ iterations, which potentially reduces the search space by many orders of magnitude if $m << k$. This algorithm is nonetheless highly inefficient, however, it is still feasible in many important cases. There are undoubtedly many performance improvements which could be made to this algorithm, but this is left as a topic for future research.

This algorithm will consistently estimate the unique valid state-space model as $n \rightarrow \infty$ with $k \equiv |\mathbf{w}|$ fixed. The test given by \citeauthor{srivastava2005some} (\citeyear{srivastava2005some}), and indeed the multiple testing strategy has asymptotic power equal to unity. Therefore, since the algorithm systematically considers every possible model, it will reject every incorrect model in the asymptotic case. It will also reject the correct model in a proportion $\alpha$ of samples. In these cases the algorithm will (still asymptotically) yield no solution. In the rest it will yield the unique valid model. 

\vspace{0.5cm}

\begin{minipage}{\linewidth}
  \begin{algorithm}[H]
    \SetAlgoLined
    \DontPrintSemicolon
    
    \SetStartEndCondition{ }{}{}%
    \SetKwProg{Fn}{def}{\string:}{}
    \SetKwFunction{Range}{range}
    \SetKw{KwTo}{in}\SetKwFor{For}{for}{\string:}{}%
    \SetKwIF{If}{ElseIf}{Else}{if}{:}{elif}{else:}{}%
    \SetKwFor{While}{while}{:}{fintq}%
    \AlgoDontDisplayBlockMarkers
    \SetAlgoNoEnd

    \SetKwData{Continue}{$continue$}
    \SetKwData{Nstates}{$n\_states$}
    \SetKwData{Maxstates}{$max\_states$}
    \SetKwData{Allvalidstates}{$all\_valid\_states$}
    \SetKwData{Siglevel}{$sig\_level$}
    \SetKwData{Potentialstates}{$potential\_states$}
    \SetKwData{Allpotentialstates}{$all\_potential\_states$}
    \SetKwData{Potentialstate}{$potential\_state$}
    \SetKwData{Constrainttests}{$constraint\_tests$}
    \SetKwData{Constrainttest}{$constraint\_test$}
    \SetKwData{Scoretests}{$score\_tests$}
    \SetKwData{Scoretest}{$score\_test$}
    \SetKwData{Nendo}{$\#endogenous\_states$}
    \SetKwData{States}{$states$}
    \SetKwData{True}{$true$}
    \SetKwData{False}{$false$}
    \SetKwFunction{Getpotentialstates}{$get\_potential\_states$}
    \SetKwData{Getconstrainttests}{$get\_constraint\_tests$}
    \SetKwData{Getscoretests}{$get\_score\_tests$}
    \KwIn{$alpha$: significance level}
    \KwIn{$test$: testing strategy is either 'multiple' or 'srivastava'}
    \KwOut{$all\_valid\_states$: A set of minimal sets of exogenous and endogenous states whose implied conditional independences are valid relative to the observed data, sorted by likelihood}
    \Begin{
      \Continue $=$ \True\;
      \Nstates $= 0$\;
      \Maxstates $= \#observables - 2$\;
      \Allvalidstates $= list()$ \;
      \While{\Continue $and$ \Nstates $<=$ \Maxstates} {
        \Allpotentialstates $=$ \Getpotentialstates{\Nstates}\;
        \For{\Potentialstates $\in$ \Allpotentialstates}{
          \Constrainttests $=$ \Getconstrainttests{\Potentialstates}\;
          \Scoretests $=$ \Getscoretests{\Potentialstates}\;
          \If{$test$ $=$ $multiple$}{
            \Siglevel $=$ $\frac{alpha}{length(constraint\_tests)}$\;
          } \Else {
            \Siglevel $=$ $alpha$
          }
          \If{$every$ \Constrainttest $.p\_value >$ \Siglevel $for$ \Constrainttest $\in$ \Constrainttests}{
            $append$ \Potentialstates $to$ \Allvalidstates\;
            \Continue $=$ \False\;
          }
        }
      }
      $sort$ $descending$ \Allvalidstates $by$ \Nendo, \Scoretests\;
      $return$ \Allvalidstates
  }
    \caption{Brute force hybrid state-space estimation algorithm}
    \label{constraint_algo}
  \end{algorithm}
\end{minipage} \\

However, in finite samples there is unfortunately no guarantee that the algorithm will yield the correct solution. Although the test from \citeauthor{srivastava2005some} (\citeyear{srivastava2005some}) is only asymptotically normal, in practice the type I error rate remains close to the specified significance level $\alpha$ for any reasonable sample size (see Appendix \ref{testing_validation}), so this seems to be a reasonable approximation. On the other hand, the probability of type II error can be quite high in small samples, and this is very problematic. The algorithm will stop early if it finds some valid model with $m$ states. However, if this is the result of a type II error, and the correct model actually has more than $m$ states, then the algorithm will terminate before it ever even considers the correct model. Potential solutions to this problem that would improve small sample performance would be to devise a test with more power, remove the early stopping behaviour, or otherwise limit the size of the search space, although this may result in a greater reliance on sorting by score to differentiate between valid models, or a loss of the agnosticism of the algorithm.

\subsection{Related Modelling Techniques}

Having discussed how DSGE models and macroeconomic data more generally can be represented as DAGs this section will discuss how this approach relates to other econometric approaches which are common in the analysis of macroeconomic time-series. It is possible to draw comparisons with both Structural Vector Autoregression (SVAR) and Autoregessive Distributed Lag (ADL) models, so these will be discussed in turn.

One of the most common and simplest econometric models for this type of data is the vector autoregression (VAR), which was introduced by \citeauthor{sims1980macroeconomics} (\citeyear{sims1980macroeconomics}). This method involves regressing a vector of outcomes $\mathbf{y}_t$ on a matrix containing $p$ lags of $\mathbf{y}_t$ in the form $\mathbf{y}_t = [\mathbf{y}_{t-1}, ..., \mathbf{y}_{t-p}] \beta + \mathbf{\epsilon}_t $. The primary concern with and limitation of this approach is that the estimated covariance matrix $\mathbf{\epsilon}_t$ is unrestricted, so the shocks contained within it are not mutually independent. Therefore, this model can not be used to estimate the effect of a truly exogenous shock on the dynamics of observed variables. In order to address this issue the model can transformed and an assumed causal ordering imposed in the form of a Cholesky decomposition \parencite{sims1980macroeconomics}, which has the effect of making the errors of the estimated, transformed model mutually uncorrelated or in other words \textit{structural}. Therefore, such models are known as SVARs. As noted by \citeauthor{demiralp2003searching} (\citeyear{demiralp2003searching}), there is an equivalence that can be drawn between SVAR models and DAGs. Indeed, in this paper they implement the PC-algorithm \parencite{spirtes1991algorithm} to show that structure learning methods for DAGs can be used to identify the causal order of the Cholesky decomposition for an SVAR from data.

However, one key difference between a DAG in this context and a SVAR model is that the DAG allows for some variables to depend on contemporaneous values of other variables. In particular, the endogenous states and controls depend contemporaneously on the exogenous states. In this sense the DAG is similar to an Autoregressive Distributed Lag (ADL) model. When implementing an ADL model it is necessary for the researcher to choose which contemporaneous variables to include as regressors, implicitly assuming that these regressors are exogenous relative to the outcomes of interest. 

The primary advantage of DAGs is the relatively weak assumptions they require. Both the SVAR and ADL models require the researcher to specify assumptions about the relative exogeneity of observable variables. These assumptions are themselves either derived from a similarly assumption-heavy model such as a DSGE model, or are in some cases entirely \textit{ad hoc}. There has been a long tradition within the field of economics including seminal papers by \citeauthor{lucas1976econometric} (\citeyear{lucas1976econometric}), \citeauthor{sims1980macroeconomics} (\citeyear{sims1980macroeconomics}), and \citeauthor{jorda2005estimation}, (\citeyear{jorda2005estimation}) criticising this type of methodology. Seen in this way, DAGs constitute a powerful new tool to choose the specification of these types of models in an agnostic and data-driven way.

\subsection{IRFs}

One very common way of evaluating DSGE models is to compare the Impulse Response Functions (IRFs) that they imply with the IRFs of reduced form models such as VAR models \parencite[p.83]{ramey2016handbook}. This is also possible when directly estimating state-space models, and the results of this will be considered in the empirical section of this paper. This is simply done to demonstrate that the state-space model that is estimated matches the reduced form of the original simulation. IRFs are calculated starting with a vector of initial values (shocks), by iteratively using the estimated matrices $\hat{\mathbf{A}}$ - $\hat{\mathbf{E}}$ to calculate current time step values using past values. Note that this can be done for either exogenous or endogenous states, but not for controls, as changes to these are by construction not propagated through to future time steps.

\section{Data} \label{data}

In order to demonstrate the capability of the proposed algorithm empirically I will work with both simulated and real macroeconomic data. Using simulated data has a number of advantages. Firstly, since the model that generates the data is known it is possible to evaluate whether structure learning has succeeded in identifying the ground-truth DAG. Secondly, in this context we can ensure to the greatest possible extent that the underlying assumptions of the structure learning algorithms, including linearity and normality are satisfied. Finally, since these models are standard in modern macroeconomics it provides a highly relevant controlled testing environment. On the other hand, using real data is an opportunity to demonstrate that the algorithm is also a powerful heuristic tool that can be implemented outside a rigorously controlled environment. Furthermore, if these results are to be believed it will allow for inferences pertaining to a number of important debates in the DSGE literature. The remainder of this section will discuss the various sources and general properties of the data used.

\subsection{Simulations}

In order to collect simulated data I consulted a GitHub repository containing Dynare code to replicate well known macroeconomic models \parencite{pfeifer2020}. In particular, I chose to model the baseline RBC model as a simple case and a New Keynesian model from \citeauthor{gali2015monetary} (\citeyear{gali2015monetary}) for a more difficult and complex modelling challenge. Simulations output a file containing $100,000$ observations of $i.i.d.$ draws of the exogenous shocks, and the associated observed values of the other variables in the model. This data and smaller subsets thereof, were then used as inputs for the structure learning algorithm.

\subsubsection{Baseline RBC}

\begin{table}
  \centering
  \begin{tabular}{|l|l|l|}
    \hline
    Symbol & Name & Type \\
    \hline
    $g$ & government spending & exogenous state \\
    $z$ & technology & exogenous state \\
    $k$ & capital & endogenous state \\
    $w$ & wage rate & control \\
    $r$ & return to capital & control \\
    $y$ & output & control \\
    $c$ & consumption & control \\
    $l$ & hours worked & control \\
    $i$ & investment & control \\ \cline{2-2}
    \hline
  \end{tabular}
  \caption{Description of variables for the baseline RBC model.}
  \label{rbc_data}
\end{table}

The baseline RBC model includes 11 variables which are summarised by Table \ref{rbc_data}. This model contains two exogenous state variables: technology and government spending, and one endogenous state: capital. There are two shocks in the model: one that affects only technology directly and one that affects only government spending directly. As explained in Section \ref{dsge}, these shocks will be dropped from the data. The shocks are Gaussian and orthogonal, and furthermore the model is taken as a first-order approximation. Therefore, all of the necessary assumptions are satisfied.

This model was chosen as it is one of the simplest DSGE models and provides a good baseline to demonstrate the effectiveness of this methodology. In particular, the default calibration of this model which was used has autoregressive coefficients on the exogenous technology and government spending processes that are very close to one, and as a result there is a high degree of persistence in all variables in the model. This model will test the algorithm's performance when the assumption of stationarity is challenged.

\subsubsection{Baseline New Keynesian}

\begin{table}
  \centering
  \begin{tabular}{|l|l|l|}
    \hline
    Symbol & Name & Type \\
    \hline
    $nu$ & policy rate & exogenous state \\
    $a$ & technology & exogenous state \\
    $z$ & preferences & exogenous state \\
    $p$ & price level & endogenous state \\
    $y$ & output & control \\
    $i$ & nominal interest & control \\
    $pi$ & inflation & control \\
    $y\_gap$ & output gap & control \\
    $r\_nat$ & natural interest rate & control \\
    $r\_real$ & real interest rate & control \\
    $n$ & hours worked & control \\
    $m\_real$ & real money balances & control \\
    $m\_nominal$ & nominal money balances & control \\
    $w$ & nominal wages & control \\ 
    $c$ & consumption & control \\
    $w\_real$ & real wages & control \\
    $mu$ & mark-up & control \\ \cline{2-2}
    \hline
  \end{tabular}
  \caption{Description of variables for the baseline New Keynesian model.}
  \label{nk_data}
\end{table}

New Keynesian models are extremely popular in modern macroeconomics and are also considerably more complex than the baseline RBC. Therefore, this serves as a worthy challenge for this methodology. In particular, I use a model from \citeauthor{gali2015monetary} (\citeyear{gali2015monetary}) as provided by \citeauthor{pfeifer2020} (\citeyear{pfeifer2020}). The variables in this model are summarised in Table \ref{nk_data}\footnote{Some control variables which were just linear functions of another variable were dropped, for example, annualised rates.}. This model has a total of four state variables: three exogenous states (policy rate, technology and, preferences) for which there is one $i.i.d.$ and Gaussian shock each, and one endogenous state (price level).

\subsection{US Data}

\begin{table}
  \centering
  \begin{tabular}{|l|l|}
    \hline
    Symbol & Name \\
    \hline
    $pi$ & CPI Inflation \\
    $rm$ & Federal Funds Rate (Return to Money) \\
    $g$ & (Real) Government Expenditure \\
    $y$ & (Real) GDP \\
    $i$ & (Real) Private Investment \\
    $w$ & Median (Real) Wage \\
    $rk$ & Return to Capital \\
    $z$ & Total Factor Productivity \\
    $u$ & Unemployment \\
    $l$ & Total Workforce \\
    $c$ & (Real) Personal Consumption \\\cline{2-2}
    \hline
  \end{tabular}
  \caption{Description of Variables for US Data}
  \label{tab3}
\end{table}

To provide an example of real macroeconomic time-series, quarterly data from the US during the period 1985-2005 were collected from FRED (\citeyear{fred2020data}) for 15 variables outlined in Table \ref{tab3}. All of the variables were detrended and demeaned by taking the residuals of an estimated first order autoregression (as opposed to an HP filter). Total factor productivity and capital stock were provided on an annual basis and were therefore interpolated quadratically. Full details of data preprocessing are available in the project repository \parencite{hall2020git}.

Since we assume a log-linear DSGE solution, we by implication assume that the data is generated from a stationary distribution with no structural breaks. This particular data set was chosen because during this timeframe that assumption is plausibly valid. In general, structural breaks are important to model correctly, however, at present incorporating these are left as an avenue for future research.

\vspace{-0.01cm}

\section{Results} \label{results}

In this section many of the properties of the proposed algorithm will be thoroughly investigated. Using simulated data allows for the possibility of many experiments to test these properties in a controlled environment. In particular, for the models under consideration two scenarios will be presented. To demonstrate asymptotic consistency, results from the algorithm for a very large number of samples (100,000) are shown. To demonstrate the finite sample properties, results from a large number of runs of the algorithm (1,000) with a relatively small and realistic sample size (100) will be provided and discussed.

\subsection{Baseline RBC}

\begin{figure}
  \centering
  \begin{subfigure}{0.8\textwidth}
    \centering
    \includegraphics[width=\linewidth]{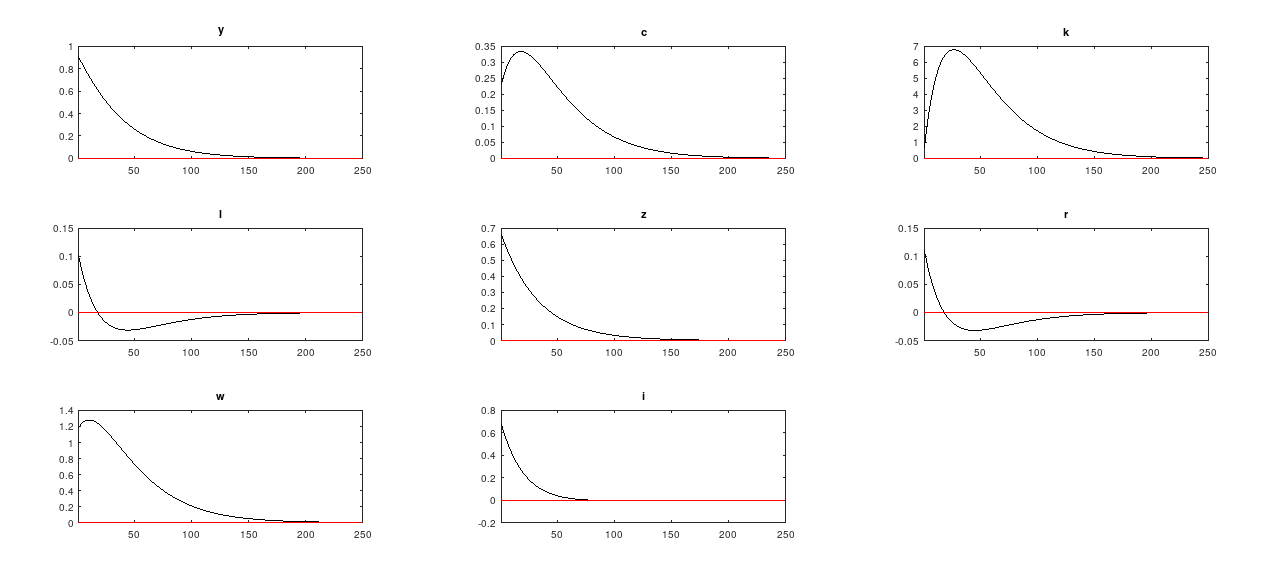} 
    \caption{Original Simulation}
    \label{rbc_simirf}
  \end{subfigure}
  \begin{subfigure}{0.8\textwidth}
    \centering  
    \includegraphics[width=\linewidth]{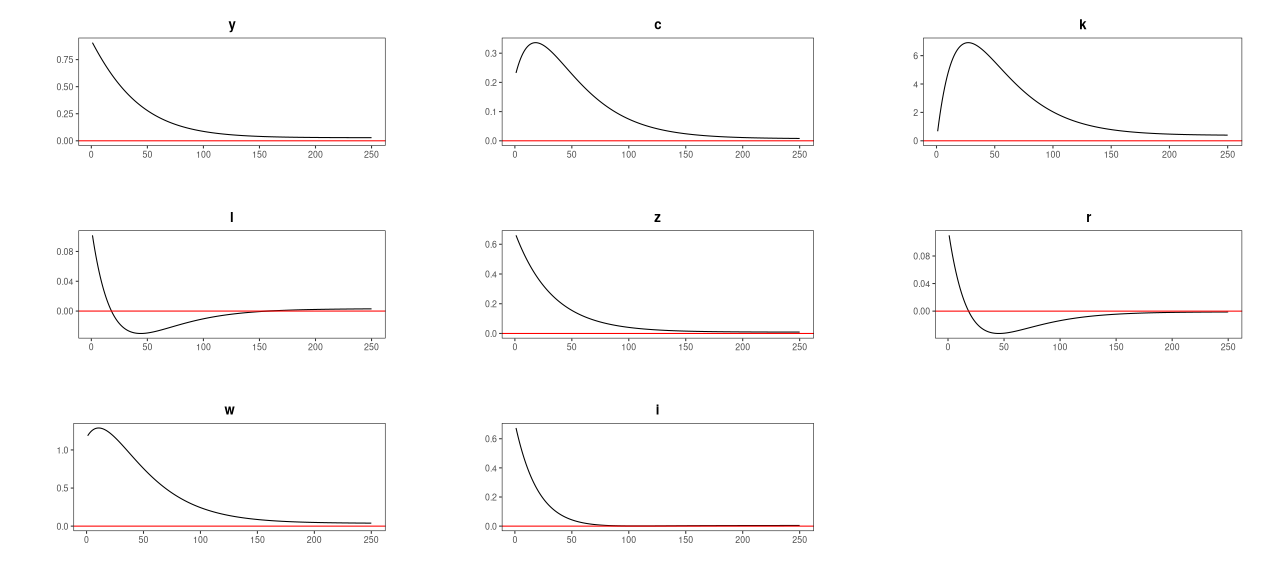}
    \caption{Estimated State-Space Model}
    \label{rbc_gtirf}
  \end{subfigure}

  \caption{IRFs to a one standard deviation technology shock generated by the original simulation and estimated model.}
  \label{rbc_irfs}
\end{figure}

Using either testing strategy from Section \ref{testing} (multiple testing or \citeauthor{srivastava2005some} (\citeyear{srivastava2005some})) on the entire sample of 100,000 observations for the RBC model the algorithm successfully identifies the correct states, which are exogenous states $z$ and $g$, and endogenous state $k$. No other (incorrect) models are valid. Figure \ref{rbc_irfs} shows the impulse responses to a technology shock generated by the original simulation and the estimated model. There are almost identical, as they should be. This is a simple validation that the selected state-space model is equivalent to the true reduced form and that we have recovered the correct parameters using maximum likelihood estimation of the associated coefficient matrices.

\begin{table}
  \centering
  \begin{tabular}{|c|c|c|l|l|}
    \bfseries Index & \bfseries Exogenous States & \bfseries Endogenous States & \bfseries Wins & \bfseries Valid
    \begin{filecontents*}{./files/rbc_wins_srivastava.csv}
index,exostates,endostates,wins,valid
1,g   z,k,944,944
2,g   w,k,27,729
3,g   y,k,27,571
4,c   g,k,2,8
5,g   l   y,,0,340
6,g   r   y,,0,421
7,g   r,k,0,576
8,g   l   z,,0,716
9,g   i   r,,0,781
10,g   i   l,,0,629
11,g   i,k,0,867
12,g   r   w,,0,609
13,g   r   z,,0,858
14,g   k   l,,0,625
15,g   l   w,,0,603
16,g   k   r,,0,779
17,c   g   w,,0,1
\end{filecontents*}
\csvreader[head to column names]{./files/rbc_wins_srivastava.csv}{}
    {\\\index & \exostates & \endostates & \wins & \valid}
  \end{tabular}
  \caption{Small-sample (n=100) simulation structure learning results for the RBC model using the \citeauthor{srivastava2005some} (\citeyear{srivastava2005some}) test. Algorithm was run for 1,000 iterations on different samples. Only models that were \textbf{Valid} relative to the conditional independence test in at least one iteration are displayed. \textbf{Wins} indicates the number of iterations in which that model was selected by the algorithm. The ground-truth model has \textbf{Exogenous States} $g$ and $z$ and \textbf{Endogenous State} $k$.}
  \label{rbcwins_sr}
\end{table}

\begin{table}
  \centering
  \begin{tabular}{|c|c|c|l|l|}
    \bfseries Index & \bfseries Exogenous States & \bfseries Endogenous States & \bfseries Wins & \bfseries Valid
    \begin{filecontents*}{./files/rbc_wins_multiple.csv}
index,exostates,endostates,wins,valid
1,g   z,k,888,997
2,c   l,k,109,109
3,g   r,k,2,941
4,g   w,k,1,986
5,g   y,k,0,974
6,g   i,k,0,996
7,c   g,k,0,200
8,g   l,k,0,4
\end{filecontents*}
\csvreader[head to column names]{./files/rbc_wins_multiple.csv}{}
    {\\\index & \exostates & \endostates & \wins & \valid}
  \end{tabular}
  \caption{Small-sample (n=100) simulation structure learning results for the RBC model using pairwise correlation tests and a \citeauthor{bonferroni1936teoria} (\citeyear{bonferroni1936teoria}) correction (multiple testing strategy). Algorithm was run for 1,000 iterations on different samples. Only models that were \textbf{Valid} relative to the conditional independence test in at least one iteration are displayed. \textbf{Wins} indicates the number of iterations in which that model was selected by the algorithm. The ground-truth model has \textbf{Exogenous States} $g$ and $z$ and \textbf{Endogenous State} $k$.}
  \label{rbcwins_mu}
\end{table}

Table \ref{rbcwins_sr} shows the small sample results for the algorithm using the test based on \citeauthor{srivastava2005some} (\citeyear{srivastava2005some}), and Table \ref{rbcwins_mu} likewise for the multiple testing strategy. We will now discuss each of these results in turn.

The results in Table \ref{rbcwins_sr} are promising for a number of reasons. The headline result is that the ground-truth model (with exogenous states $g$ and $z$ and endogenous state $k$) is selected by the algorithm (denoted as "wins") in nearly 95\% of iterations, and in every iteration where it is valid. The latter observation suggests that sorting by number of endogenous states and the likelihood function is having the intended effect. Also note that the empirical size of the test is quite close to the expected 5\% significance level, as the correct model was rejected 56 times out of 1,000 iterations ($\sim 5.6\%$). We can see that out of the 834 models that are considered in each iteration, that is, the models with less than or equal to three state variables, only 17 ($\sim 2\%$) are ever valid, and of those 17 only 4 (including the true model) are ever selected as the optimal model by the algorithm. Therefore, this testing strategy seems to have strong power to reject incorrect models in this application. 

Table \ref{rbcwins_mu} mirrors the previous results in many ways, however, there are some key differences. The correct model is only rejected in 3 out of the 1,000 iterations, so the empirical size is far below the specified 5\% significance level. This confirms suspicions that these pairwise correlation tests are not independent. However, this is much better than having higher than expected type I error, and this low type I error rate does not seem to have come at the cost of power, at least in comparison to the other testing strategy. Here only 8 out of the 834 models considered were ever valid, so this testing strategy actually seems to have higher power. Nonetheless, the true model does win less often using this approach (only 888 times as compared with 944), primarily because the model with exogenous states $c$ and $l$ wins 109 times (every time it is valid). This particular model was rejected in every iteration of the \citeauthor{srivastava2005some} (\citeyear{srivastava2005some}) test, despite its overall lower power. It seems likely that this particular combination of states performs so well in the multiple testing approach because these $c$ and $l$ are nearly collinear with the true exogenous states $z$ and $g$, while being even more persistent, with very high estimated autoregressive coefficients of $0.994$ and $0.972$ respectively. As a result the prediction while treating $g$ and $z$ as controls obtains a relatively high likelihood score. The conclusion here is that this approach may run into difficulties in small samples if there is a very high degree of multicollinearity or autocorrelation among observables. 

\subsection{Baseline New Keynesian} \label{nk_results}

We now turn our attention to the more complex baseline New Keynesian model. This model contains 17 observables, and is thus considerably more complex than the simulated RBC data. Table \ref{nk_full_mu} shows the results for a large sample, using the multiple testing approach. The Srivastava approach are not shown, because this approach did not work in this application. This test is lacking in power (even with the full sample), such that numerous models with only two states were found to be valid, and therefore the algorithm terminated before considering the ground truth, which has four states. This unfortunately highlights one of the limitations to this approach.

On the other hand, while using the multiple testing strategy, the results are still promising. While using the full sample of $100,000$ observations, only two models are valid, and the correct model with exogenous states $a$, $nu$, and, $z$ and endogenous state $p$ wins both on preference for models with more endogenous states and on log-likelihood. This once again constitutes empirical validation of asymptotic properties.

\begin{table}
  \centering
  \begin{tabular}{|c|c|c|c|}
    \bfseries Index & \bfseries Exogenous States & \bfseries Endogenous States &  \bfseries Log-Likelihood
    \begin{filecontents*}{./files/nk_full_multi.csv}
index,exostates,endostates,loglik
1,a   nu   z,p,28426392.79
2,a   nu   z   p,,27640572.98
\end{filecontents*}
\csvreader[head to column names]{./files/nk_full_multi.csv}{}
    {\\\index & \exostates & \endostates & \loglik}
  \end{tabular}
  \caption{Large sample (n=100,000) simulation structure learning results for one run using the New Keynesian model using pairwise correlation tests and a \citeauthor{bonferroni1936teoria} (\citeyear{bonferroni1936teoria}) correction. The ground-truth model has \textbf{Exogenous States} $a$, $nu$ and, $z$ and \textbf{Endogenous State} $p$.}
  \label{nk_full_mu}
\end{table}

\begin{table}
  \centering
  \begin{tabular}{|c|c|c|l|l|}
    \bfseries Index & \bfseries Exogenous States & \bfseries Endogenous States &  \bfseries Wins & \bfseries Valid
    \begin{filecontents*}{./files/nk_wins_multiple.csv}
index,exostates,endostates,wins,valid
1,a   nu   z,p,753,999
2,a   mu   nu,p,90,183
3,a   nu   w   real,p,72,217
4,m   real   nu   z,p,20,670
5,m   real   nu   r   nat,p,18,63
6,a   nu   r   nat,p,18,966
7,nu   pi   y,p,14,375
8,a   r   real   w   real,p,4,6
9,a   nu   p   z,,3,1000
10,a   mu   nu   p,,2,181
11,mu   nu   y,p,2,240
12,a   n   nu,p,1,183
13,a   mu   r   real,p,1,2
14,nu   y   z,p,1,26
15,i   nu   y,p,1,742
\end{filecontents*}
\csvreader[head to column names]{./files/nk_wins_multiple.csv}{}
    {\\\index & \exostates & \endostates & \wins & \valid}
  \end{tabular}
  \caption{Small-sample (n=100) simulation structure learning results for New Keynesian model using pairwise correlation tests and a \citeauthor{bonferroni1936teoria} (\citeyear{bonferroni1936teoria}) correction. Algorithm was run for 1,000 iterations on different samples. Only models that had at least one \textbf{Win} are displayed. 55 models were \textbf{Valid} relative to the conditional independence test in at least one iteration. \textbf{Wins} indicates the number of iterations in which that model was chosen as optimal by the algorithm. The ground-truth model has \textbf{Exogenous States} $a$, $nu$, and $z$ and \textbf{Endogenous State} $p$.}
  \label{nkwins_mu}
\end{table}

Table \ref{nkwins_mu} shows the small sample results using the multiple testing strategy. The results are not as strong as with the RBC model, but this is to be expected given the greater number of variables and complexity of model considered with the same sample size. We find that the ground-truth model wins in approximately $75\%$ of iterations, while only being rejected once. Fifty-five models were valid in at least one iteration, which represents approximately $0.1\%$ of models tested in each iteration. So despite the complexity of this problem, the actual type I and type II error rates of the multiple testing strategy in this application were actually even lower in relative terms than in the RBC setup, only not by enough to completely offset the increased complexity of the problem. Compared to that setup there were more valid models in any given iteration. As a result the task of sorting left over models by likelihood is more difficult, and this explains why the true model is not chosen as often, despite almost always being valid.

These results show that there are practical limitations to how well the algorithm and tests can perform. The tests are consistent as the sample size $n \rightarrow \infty$ with the number of observables $k$ fixed. If $k$ is not so small compared to the sample size then there is likely to be poor performance. This is a problem common to all high-dimensional econometric models, however, it may be particularly acute here because the number of models considered, and thus the complexity of the problem grows exponentially in the number of observables.

\subsection{US Data}

\begin{table}
  \centering
  \begin{tabular}{|c|c|c|c|}
    \bfseries Index & \bfseries Exogenous States & \bfseries Endogenous States & \bfseries Log-Likelihood
    \begin{filecontents*}{./files/realresults_multiple.csv}
index,exostates,endostates,controls,loglik
1,y z u,pi rm k c,g i w rk l,3759.52
2,rm y z u,pi k c,g i w rk l,3743.85

\end{filecontents*}
\csvreader[head to column names]{./files/realresults_multiple.csv}{}
    {\\\index & \exostates & \endostates & \loglik}
  \end{tabular}
  \caption{Structure learning results for the US macroeconomic data set (1985-2005) using pairwise correlation tests and a \citeauthor{bonferroni1936teoria} (\citeyear{bonferroni1936teoria}) correction.}
  \label{real_mu}
\end{table}

Table \ref{real_mu} shows results for structure learning on the US macroeconomic data set using the multiple testing strategy\footnote{Again, Srivastava results are not shown because, given the result from Section \ref{nk_results}, the multiple testing strategy results seem more credible.}. Despite the small data set of only 80 observations these tests were able to reject all but 2 of the $93,434$ models considered. Many features of this solution are consistent with what standard intuitions would imply. For example, we observe that capital and the policy rate are endogenous states. Both of these are standard features of any DSGE model, and the second one reflects the well documented \citeauthor{taylor1993discretion} (\citeyear{taylor1993discretion}) rule. Also note that TFP is exogenous, which is fairly standard outside endogenous growth models.

If we are to believe these results, then there are numerous implications for theory, at least in the context of US macroeconomic trends. First of all, the fact that consumption is an endogenous state is evidence in favour of the hypotheses of \citeauthor{fuhrer2000habit} (\citeyear{fuhrer2000habit}) that DSGE models should take into account habits in consumption, thus making consumption inertial. Furthermore, we observe that inflation is an endogenous state. This is evidence related to a particularly heated debate surrounding whether inflation is purely rational and forward-looking \parencite{levin2004macroeconomic}, and should therefore be modelled as a control variable, or whether inflation demonstrates persistence \parencite{christiano2005nominal} due perhaps to indexing or other forms of bounded rationality, and should therefore be modelled as a state variable. Clearly then, this evidence supports the latter hypothesis. 

Perhaps more difficult to reason about is why output and unemployment enter as exogenous states. But for these too, some explanation can be suggested. Exogenous states are the only variables in the model which are directly exposed to shocks. Recall that these shocks are assumed to be structural or orthogonal. Assuming a Cobb-Douglas style production function the three determinants of output are TFP, labour input, and the capital input. Since unemployment (which is inversely proportional to the labour input) and TFP are already included as exogenous states, orthogonal shocks to output must be shocks to the capital input. Yet capital itself is included in the observables here, therefore, this is best interpreted as a shock to variable capacity utilisation \parencite{driver2000capacity}. Similarly, unemployment may be subject to orthogonal labour market or other policy shocks. The fact that these variables enter as exogenous states suggests that these shocks are the most important in explaining the dynamics of the macroeconomy.

Figures \ref{us_zu_irfs} and \ref{us_yrm_irfs} in the appendix show IRFs for shocks to the three exogenous states, as well as to the policy rate $rm$, which is an endogenous state in the estimated state-space model and in an unconditional VAR(1) model. The VAR IRFs are provided as a basic sanity check --- so that some form of comparison can be made, but there is little reason to believe that these are necessarily a good depiction of reality. Ultimately, there is no fundamental ground-truth to compare these IRFs to, much like the choice of state variables they can only be evaluated against common heuristics and \textit{stylised facts} in the literature. For example, consider the response to a TFP shock. All the IRFs from the state-space model match the direction of those from the VAR, except output and unemployment (which are exogenous and thus do not respond in the state-space model), and labour force, which responds negatively in the state-space IRFs. Regarding the last point however, the state-space model is probably more credible than the VAR, as declining labour input as a response to technology shocks is a well documented empirical fact \parencite{gali2004technology}. Now consider the IRFs generated for a (expansionary) monetary policy shock. Again, the state-space IRFs match the direction of those from the VAR for the non-exogenous variables, except for investment, which is markedly different. The state-space predicts an expansion in investment after the monetary policy shock, whereas the VAR predicts a decrease. Again, the state-space model seems to be in agreement with empirical work in this area such as that of \citeauthor{christiano2005nominal} (\citeyear{christiano2005nominal}).

The purpose of this exercise is not to argue that this is the optimal model for macroeconomic behaviour in the United States, but rather, it is to demonstrate that the algorithm provides sensible results when used outside the laboratory setting provided by the simulated data used in previous sections. Hopefully having convinced the reader of this, I will forgo any deeper analysis of the IRFs produced by this model. It is entirely possible to use this approach to estimate a model that is worthy of such further discussion, perhaps even to go so far as specifying a micro-founded model that is consistent with the conclusions of the algorithm, but this is for now left as an avenue for future research.

\subsection{Alternative Approaches}

This section will briefly present some other strategies suggested by the literature, which I found to be less successful in this application, and will briefly discuss some reasons why that was the case. For ease of comparison only results for the RBC model are shown for each approach.

\subsubsection{Score Maximisation} \label{score}

The main approach implements a \textit{score function} to differentiate between models only when more than one survives the conditional independence tests. As discussed in Section \ref{dags}, it is at least theoretically possible to learn the structure of a DAG using only the score function. In order to implement this we attempt to maximise, in a brute-force fashion, the BIC score \parencite{schwarz1978estimating} and the log-likelihood over the set of all possible state-space models, which as discussed in Section \ref{methodology} is a subset all possible DAGs. Tables \ref{rbcwins_bic} and \ref{rbcwins_ll} respectively display full sample (n=100,000) results using these two different score functions.  

\begin{table}
  \centering
  \begin{tabular}{|c|c|c|l|}
    \bfseries Index & \bfseries Exogenous States & \bfseries Endogenous States & \bfseries BIC
    \begin{filecontents*}{./files/rbc_full_bic.csv}
index,endostates,exostates,bic
1,r,,4297163.84
2,,r,4279591.02
3,l,,4136639.26
4,g,,4121708.70
5,,l,4118845.19
6,,g,4115421.56
7,i,,3824169.94
8,l,r,3806905.97
9,r,l,3781448.32
10,,i,3754446.51
\end{filecontents*}
\csvreader[head to column names]{./files/rbc_full_bic.csv}{}
    {\\\index & \endostates & \exostates & \bic}
  \end{tabular}
  \caption{
    Large sample (n=100,000) simulation structure learning results for RBC data over one run by maximising the BIC score function only. The top 10 models are shown. Since there are no conditional independence tests, all models are \textbf{Valid}. The ground-truth model has \textbf{Exogenous States} $g$ and $z$ and \textbf{Endogenous State} $k$. Using this approach the ground truth was ranked 6353 out of 16866 models.}
  \label{rbcwins_bic}
\end{table}

\begin{table}
  \centering
  \begin{tabular}{|c|c|c|l|}
    \bfseries Index & \bfseries Exogenous States & \bfseries Endogenous States & \bfseries Log-Likelihood
    \begin{filecontents*}{./files/rbc_full_ll.csv}
index,endostates,exostates,loglik
1,,k   g   r,19490138.70
2,y   c,k   z   g,19477436.32
3,y   c   z   r,k   g   i,19459969.45
4,y   l   r   i,k   z   g,19450590.03
5,i,k   z   g,19448489.62
6,c   l   z,k   g   i,19447374.64
7,y   c   l   i,k   z   g,19438489.22
8,c   z   r   w,k   g   i,19424714.81
9,y   c   l   r,k   z   g,19416553.82
10,y   c   r   i,k   z   g,19411841.94

\end{filecontents*}
\csvreader[head to column names]{./files/rbc_full_ll.csv}{}
    {\\\index & \endostates & \exostates & \loglik}
  \end{tabular}
  \caption{
    Large sample (n=100,000) simulation structure learning results for RBC data over one run by maximising the log-likelihood function only. The top 10 models are shown. Since there are no conditional independence tests all models are \textbf{Valid}. The ground-truth model has \textbf{Exogenous States} $g$ and $z$ and \textbf{Endogenous State} $k$. Using this approach the ground truth was ranked 10606 out of 16866 models.}
  \label{rbcwins_ll}
\end{table}

The results are much weaker compared to the preferred hybrid approach with conditional independence testing. The ground-truth model is obtains a very low rank using both scores, although the BIC seems to perform somewhat better, ranking the ground truth at 6353 which is higher than rank 10606 from the log-likelihood. Now consider the top 10 ranked models generated by each score function, as displayed by Tables \ref{rbcwins_bic} and \ref{rbcwins_ll}. We can see the bias towards complexity from maximising the log-likelihood, where all but the first ranked model have too many state variables. On the other hand, we can see how the BIC has regulated the complexity of the chosen models, but has done so too aggressively, and chosen models with too few state variables. The top 10 for the AIC score was the same as for the BIC and is therefore not shown. These results show that while sorting by score may be helpful for selection among already similar models, it does not seem to be very effective when used as the only model selection criteria in this application.

\subsubsection{2-phase Restricted Maximisation (Hybrid Algorithm)} \label{rsmax2}

\begin{figure}
  \centering  
  \includegraphics[width=0.6\linewidth]{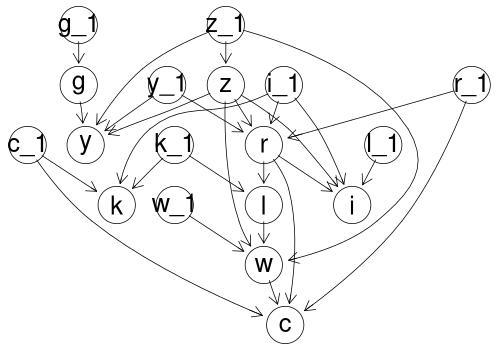}
  \caption{DAG fit to RBC data using rsmax2 hybrid constraint-based algorithm \parencite{scutari2014multiple}. Additional constraint was added such that lagged values were forced to be root nodes (as they are in the ground truth).}
  \label{hdag}
\end{figure}

\begin{figure}
  \centering  
  \includegraphics[width=0.6\linewidth]{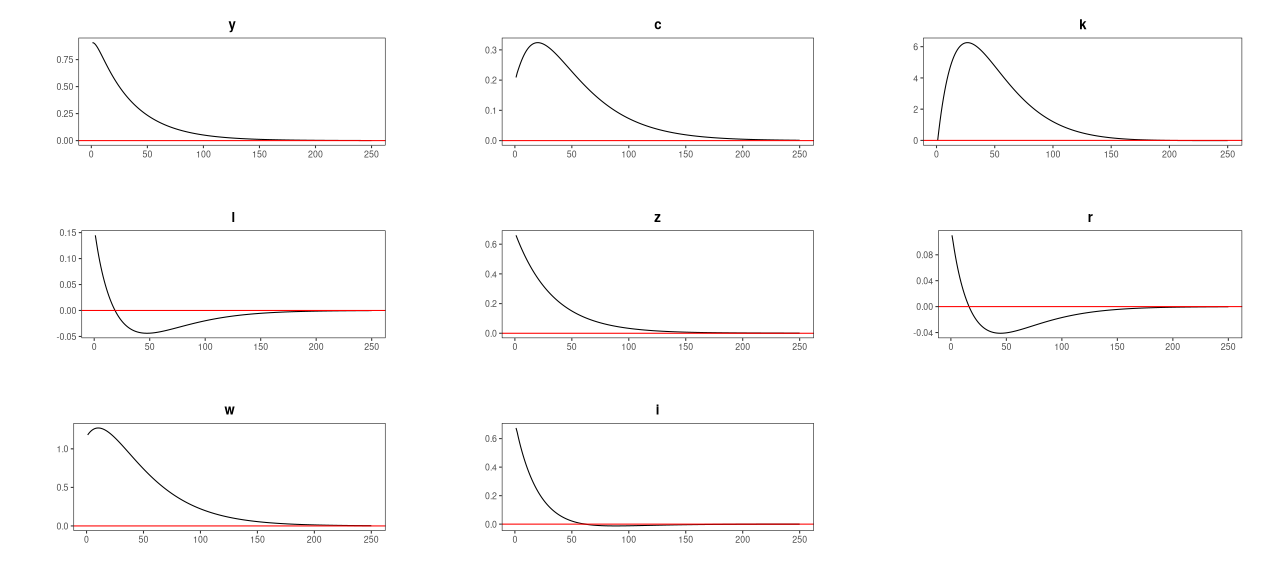}
  \caption{IRFs generated by DAG fit to RBC data using rsmax2 hybrid constraint-based algorithm \parencite{scutari2014multiple}.}
  \label{hirf}
\end{figure}

We now turn to existing structure learning methods such as the hybrid rsmax2 algorithm \parencite{scutari2014multiple}. The DAG estimated using this algorithm on a large sample of $100,000$ observations is shown in Figure \ref{hdag}. The primary limitation of this approach is that the algorithm will search over the space of all possible DAGs. This is clear when considering the estimated DAG, which does not conform to equations (\ref{ss_solution:x}) - (\ref{ss_solution:z}). In other words, the estimated solution must be incorrect because it is not a state-space model. That said, there is some extent to which important characteristics of the ground-truth solution to the RBC model can be seen here. For example, the exogenous states $z$ and $g$ depend only on their own lag. The IRFs produced by this DAG (shown in Figure \ref{hirf}) also seem to be very close to those of the original simulation. Therefore, we can conclude that although this approach did not yield the correct solution, it did recover some sense of causality from the underlying DGP.

\subsubsection{PC-algorithm (Constraint-Based Algorithm)} \label{pc}

\begin{figure}
  \centering
  \includegraphics[width=0.8\linewidth]{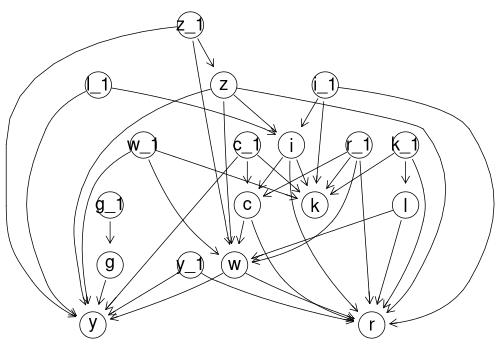}
  \caption{DAG fit to RBC data using PC-algorithm \parencite{kalisch2007estimating}. Additional constraint was added such that lagged values were forced to be root nodes (as they are in the ground truth).}
  \label{pcdag}
\end{figure}

\begin{figure}
  \centering
  \includegraphics[width=0.8\linewidth]{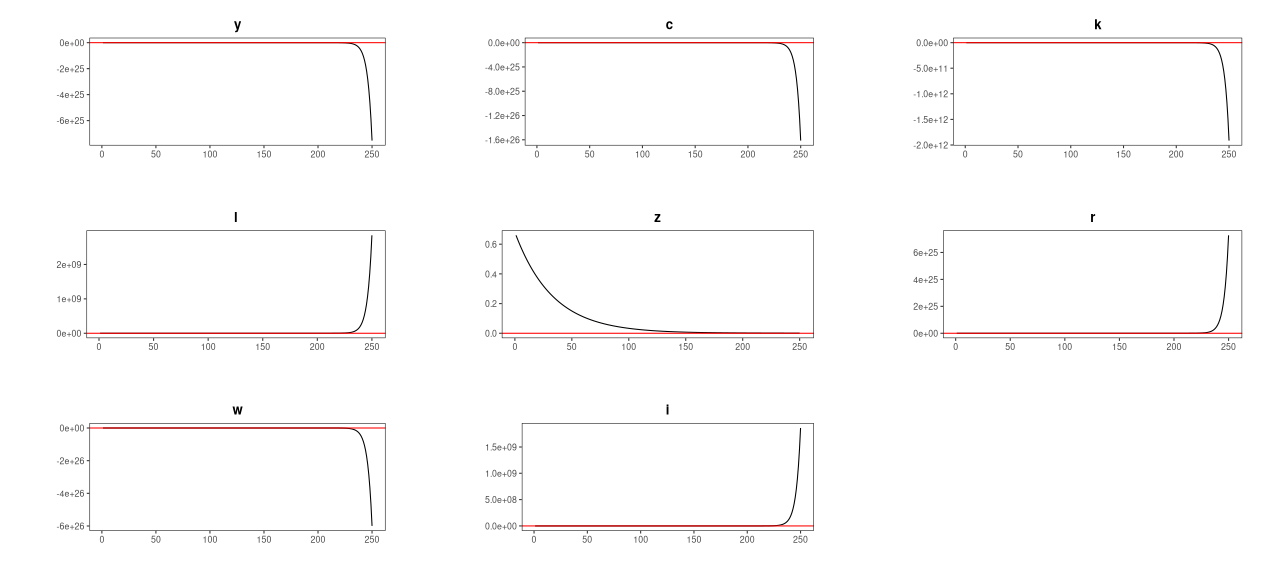}
  \caption{IRFs generated by DAG fit to RBC data using PC-algorithm \parencite{kalisch2007estimating}.}
  \label{pcirf}
\end{figure}

Finally, we consider the PC constraint-based structure learning algorithm \parencite{spirtes2000causation} \parencite{kalisch2007estimating}. The large sample estimated DAG is shown in Figure \ref{pcdag}. The result and conclusion here mirror those for the hybrid algorithm in many ways, however, this approach is somewhat less successful. In particular, the generated IRFs (Figure \ref{pcirf}) show the time-series diverging because the estimated solution is non-stationary.  

\section{Conclusion} \label{conclusion}

This paper has introduced a series of tests and an algorithm for data-driven causal discovery of macroeconomic state-space models. These tests are asymptotically consistent, and have been shown to perform well on at least relatively simple data sets given a realistic sample size. Results derived using this strategy can be used to gain insight into prominent debates in the DSGE literature. This result constitutes a concrete example of an application in which DAGs and the causal discovery toolkit more broadly can be used in empirical economics. This approach comes with a number of benefits, chief among them that it is makes no assumptions about which particular relationships are present in the ground-truth DSGE model, a property that I refer to as agnosticism.

Much work remains to be done however, as this study has uncovered a number of limitations. In order to model data from more general settings it will be necessary to incorporate DGPs that are non-stationary or contain structural breaks. Extensions could be made to allow for non-parametric conditional probability functions. These extensions are likely to come at the expense of power, however, as we have seen, the possibility of Type II error can be problematic, especially when considering complex data sets over small sample sizes. To this end new testing strategies should be developed and applied which have greater power against alternatives. Another relatively straightforward yet valuable extension to this paper would be to identify micro-founded DSGE models to match the reduced forms identified by the algorithm over some real data sets, and comparing these to the state of the art in the literature in order to see how closely they match. Finally, I have attributed the relatively poor performance of existing structure learning algorithms in this context to the large search space within which they operate. This issue seems unlikely to be unique to this particular application, suggesting that it would improve the usefulness of the implementations of these algorithms if they allowed the user to reduce the size of the search space by specifying more general constraints on the nature of the desired solution.

\citeauthor{imbens2019potential} (\citeyear{imbens2019potential}) states that DAGs are most useful, "in complex models with many variables that are not particularly popular in empirical economics." The implication is that there is limited scope for the application of (algorithmic) causal discovery in economics. However, the converse of this could equally be true; complex models with many variables are not popular in empirical economics \textit{because} there is a lack of tools such as DAGs that make these problems tractable. Particularly in the context of macroeconomics it seems that complex models such as that of \citeauthor{smets2007shocks} (\citeyear{smets2007shocks}) are becoming increasingly popular. This paper has presented an example of how the causal discovery toolkit can be used to bring data and computational power to bear on these kinds of unwieldy problems in order to derive an interpretable solution.

\newpage
\printbibliography
\newpage

\appendix

\section{Proof of Theorem \ref{validity}} \label{validity_proof}

\begin{proof}
  Suppose not. Then $M$ is faithfully represented by a DAG $H$ which is different to $G$. Since $M$ is still a log-linear DSGE solution, it must still have a faithful DAG representation of the general form in figure (\ref{dsge_dag}). Therefore, the difference must be that $H$ partitions one or more of the variables $a$ in $\mathbf{w}$ differently than $G$. Define the following notation: $G_x$ is the set of variables that are categorised as endogenous states in DAG  $G$ and likewise for $H$ and the other variable types $y$ and $z$. \\
  Consider all possible cases to see that $H$ must produce a contradiction: \\
  Case 1: $a \in G_y \text{ and } a \in H_x$ \\
    $G$ has fewer state variables that $h$, which therefore does not satisfy the MSV criteria. Contradiction. \\
  Case 2: $a \in G_y \text{ and } a \in H_z$ \\
    (\ref{constraint_test:3}) fails because there is a direct path from $\mathbf{x}_{t-1}$ to $a$ in $G$. Contradiction. \\
  Case 3: $a \in G_x \text{ and } a \in H_y$ \\
    (\ref{constraint_test:1}) fails because $a$ is not in the conditioning set for this test in $H$ and therefore there is an unblocked backdoor path from $a$ to the other time $t$ endogenous variables in $G$. Contradiction. \\
  Case 4: $a \in G_x \text{ and } a \in H_z$ \\
    (\ref{constraint_test:3}) fails because there is a direct path from $\mathbf{x}_{t-1}$ to $a$ in $G$. Contradiction. \\
  Case 5: $a \in G_z \text{ and } a \in H_y$ \\
    (\ref{constraint_test:1}) fails because there is a direct path from $a$ to any time $t$ endogenous variable in $G$. Contradiction. \\
  Case 6: $a \in G_z \text{ and } a \in H_x$ \\
    (\ref{constraint_test:1}) fails because there is a direct path from $a$ to any time $t$ endogenous variable in $G$. Contradiction. 
\end{proof}

\section{Testing Validation} \label{testing_validation}

\begin{table}[ht]
  \centering
  \begin{tabular}{|c|c|c|c|c|c|}
    \bfseries empirical Size & \bfseries Alpha & \bfseries Difference & \bfseries n & \bfseries m & \bfseries Repetitions
    \begin{filecontents*}{./files/test_validation_alpha.csv}
es,a,d,n,c,m,k
0.041,0.010,-0.031,10,0.000,5,1000
0.104,0.050,-0.054,10,0.000,5,1000
0.015,0.010,-0.005,100,0.000,5,1000
0.052,0.050,-0.002,100,0.000,5,1000
0.021,0.010,-0.011,10000,0.000,5,1000
0.031,0.050,0.019,10000,0.000,5,1000
0.118,0.010,-0.108,10,0.000,25,1000
0.239,0.050,-0.189,10,0.000,25,1000
0.012,0.010,-0.002,100,0.000,25,1000
0.041,0.050,0.009,100,0.000,25,1000
0.019,0.010,-0.009,10000,0.000,25,1000
0.053,0.050,-0.003,10000,0.000,25,1000
0.463,0.010,-0.453,10,0.000,50,1000
0.699,0.050,-0.649,10,0.000,50,1000
0.008,0.010,0.002,100,0.000,50,1000
0.076,0.050,-0.026,100,0.000,50,1000
0.008,0.010,0.002,10000,0.000,50,1000
0.063,0.050,-0.013,10000,0.000,50,1000
\end{filecontents*}
\csvreader[head to column names]{./files/test_validation_alpha.csv}{}
    {\\\es & \a & \d & \n & \m & \k}
  \end{tabular}
  \caption{empirical validation of significance level of \citeauthor{srivastava2005some} (\citeyear{srivastava2005some}) test.}
  \label{test_validation_alpha}
\end{table}

\begin{table}
  \centering
  \tiny
  \begin{tabular}{|c|c|c|c|c|c|}
    \bfseries empirical Power & \bfseries Alpha & \bfseries n & \bfseries Correlation & \bfseries m & \bfseries Repetitions
    \begin{filecontents*}{./files/test_validation_power.csv}
ep,a,n,c,m,k
0.077,0.010,10,0.100,5,1000
0.110,0.050,10,0.100,5,1000
0.379,0.010,100,0.100,5,1000
0.507,0.050,100,0.100,5,1000
1.000,0.010,10000,0.100,5,1000
1.000,0.050,10000,0.100,5,1000
0.422,0.010,10,0.100,25,1000
0.553,0.050,10,0.100,25,1000
0.999,0.010,100,0.100,25,1000
1.000,0.050,100,0.100,25,1000
1.000,0.010,10000,0.100,25,1000
1.000,0.050,10000,0.100,25,1000
0.824,0.010,10,0.100,50,1000
0.909,0.050,10,0.100,50,1000
1.000,0.010,100,0.100,50,1000
1.000,0.050,100,0.100,50,1000
1.000,0.010,10000,0.100,50,1000
1.000,0.050,10000,0.100,50,1000
0.436,0.010,10,0.325,5,1000
0.518,0.050,10,0.325,5,1000
1.000,0.010,100,0.325,5,1000
1.000,0.050,100,0.325,5,1000
1.000,0.010,10000,0.325,5,1000
1.000,0.050,10000,0.325,5,1000
0.953,0.010,10,0.325,25,1000
0.969,0.050,10,0.325,25,1000
1.000,0.010,100,0.325,25,1000
1.000,0.050,100,0.325,25,1000
1.000,0.010,10000,0.325,25,1000
1.000,0.050,10000,0.325,25,1000
0.996,0.010,10,0.325,50,1000
0.999,0.050,10,0.325,50,1000
1.000,0.010,100,0.325,50,1000
1.000,0.050,100,0.325,50,1000
1.000,0.010,10000,0.325,50,1000
1.000,0.050,10000,0.325,50,1000
0.863,0.010,10,0.550,5,1000
0.871,0.050,10,0.550,5,1000
1.000,0.010,100,0.550,5,1000
1.000,0.050,100,0.550,5,1000
1.000,0.010,10000,0.550,5,1000
1.000,0.050,10000,0.550,5,1000
0.997,0.010,10,0.550,25,1000
1.000,0.050,10,0.550,25,1000
1.000,0.010,100,0.550,25,1000
1.000,0.050,100,0.550,25,1000
1.000,0.010,10000,0.550,25,1000
1.000,0.050,10000,0.550,25,1000
1.000,0.010,10,0.550,50,1000
1.000,0.050,10,0.550,50,1000
1.000,0.010,100,0.550,50,1000
1.000,0.050,100,0.550,50,1000
1.000,0.010,10000,0.550,50,1000
1.000,0.050,10000,0.550,50,1000
0.996,0.010,10,0.775,5,1000
0.997,0.050,10,0.775,5,1000
1.000,0.010,100,0.775,5,1000
1.000,0.050,100,0.775,5,1000
1.000,0.010,10000,0.775,5,1000
1.000,0.050,10000,0.775,5,1000
1.000,0.010,10,0.775,25,1000
1.000,0.050,10,0.775,25,1000
1.000,0.010,100,0.775,25,1000
1.000,0.050,100,0.775,25,1000
1.000,0.010,10000,0.775,25,1000
1.000,0.050,10000,0.775,25,1000
1.000,0.010,10,0.775,50,1000
1.000,0.050,10,0.775,50,1000
1.000,0.010,100,0.775,50,1000
1.000,0.050,100,0.775,50,1000
1.000,0.010,10000,0.775,50,1000
1.000,0.050,10000,0.775,50,1000
1.000,0.010,10,1.000,5,1000
1.000,0.050,10,1.000,5,1000
1.000,0.010,100,1.000,5,1000
1.000,0.050,100,1.000,5,1000
1.000,0.010,10000,1.000,5,1000
1.000,0.050,10000,1.000,5,1000
1.000,0.010,10,1.000,25,1000
1.000,0.050,10,1.000,25,1000
1.000,0.010,100,1.000,25,1000
1.000,0.050,100,1.000,25,1000
1.000,0.010,10000,1.000,25,1000
1.000,0.050,10000,1.000,25,1000
1.000,0.010,10,1.000,50,1000
1.000,0.050,10,1.000,50,1000
1.000,0.010,100,1.000,50,1000
1.000,0.050,100,1.000,50,1000
1.000,0.010,10000,1.000,50,1000
1.000,0.050,10000,1.000,50,1000
\end{filecontents*}
\csvreader[head to column names]{./files/test_validation_power.csv}{}
    {\\\ep & \a & \n & \c & \m & \k}
  \end{tabular}
  \caption{empirical validation of power of \citeauthor{srivastava2005some} (\citeyear{srivastava2005some}) test against data generated from a normal distribution where the off-diagonal elements of the covariance matrix all take on the value specified by \textit{correlation}.}
  \label{test_validation_power}
\end{table}

\FloatBarrier

\section{Real Data IRFs} \label{real_irfs}

\begin{figure}[!ht]
  \centering
  \begin{subfigure}{0.8\textwidth}
    \centering
    \includegraphics[width=\linewidth]{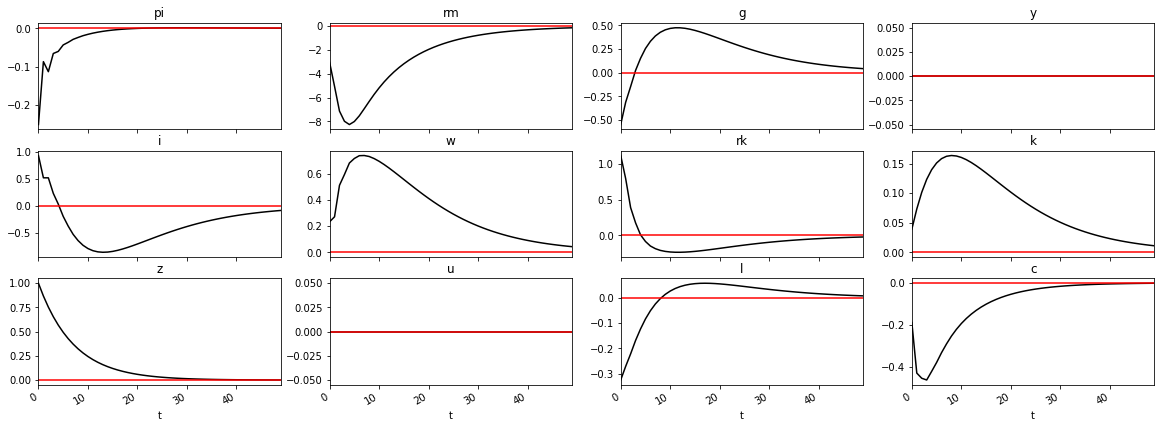} 
    \caption{TFP (State-Space)}
    \label{us_z_irf}
  \end{subfigure}
  \begin{subfigure}{0.8\textwidth}
    \centering
    \includegraphics[width=\linewidth]{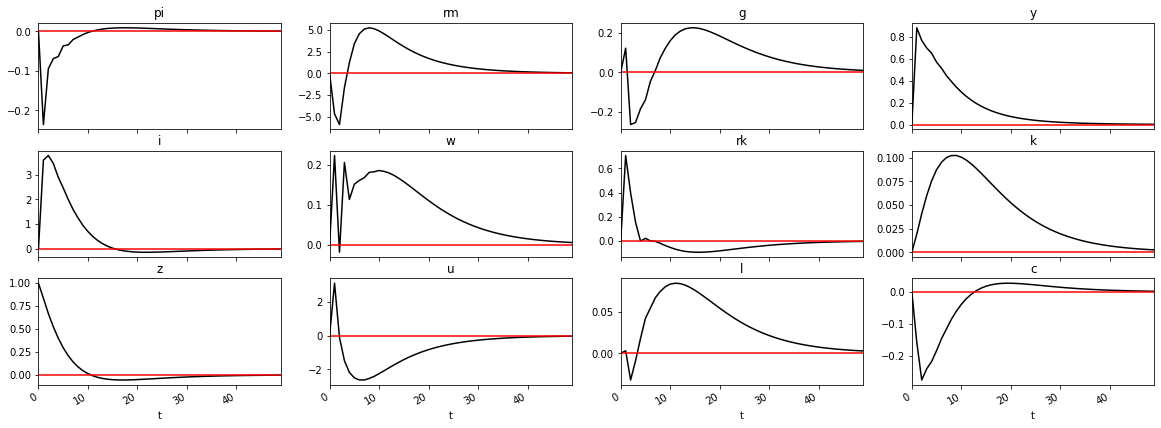} 
    \caption{TFP (VAR)}
    \label{us_z_var_irf}
  \end{subfigure}
  \begin{subfigure}{0.8\textwidth}
    \includegraphics[width=\linewidth]{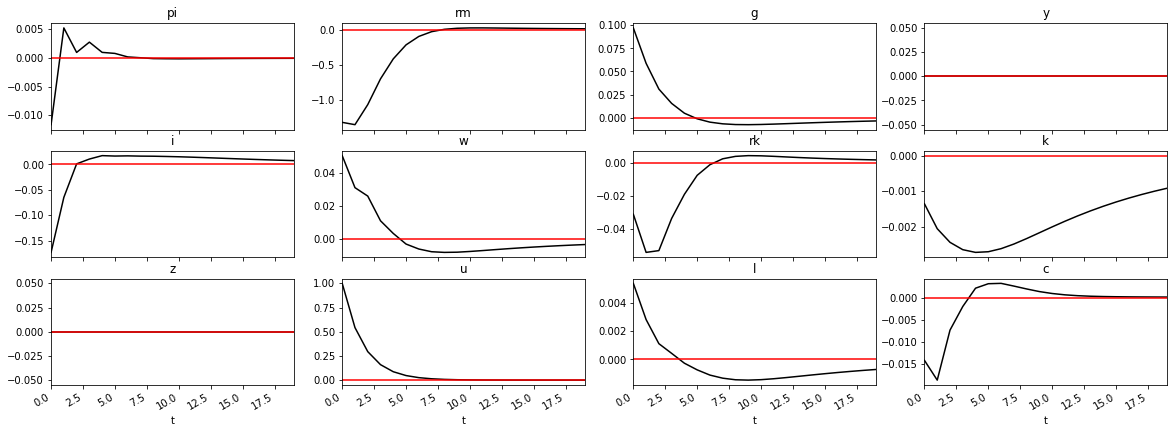}
    \caption{Unemployment (State-Space)}
    \label{us_u_irf}
  \end{subfigure}
  \begin{subfigure}{0.8\textwidth}
    \includegraphics[width=\linewidth]{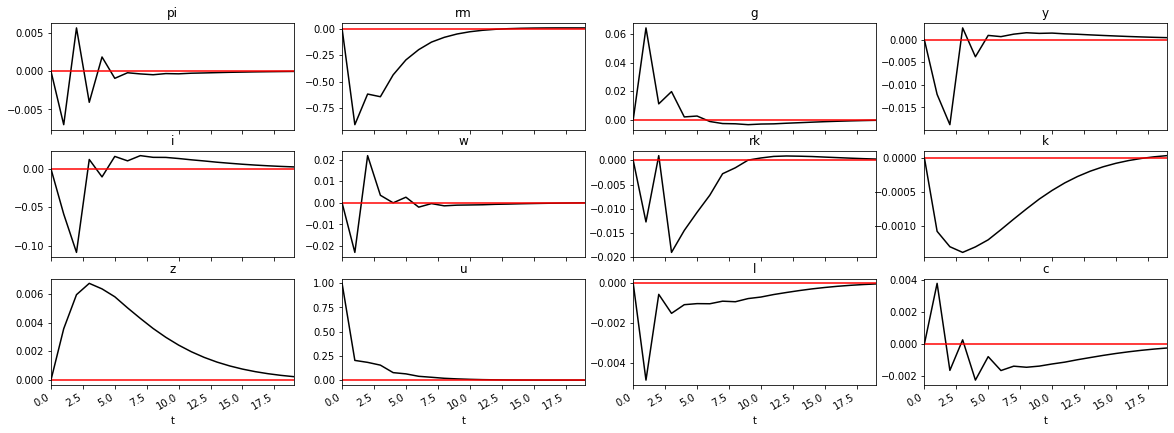}
    \caption{Unemployment (VAR)}
    \label{us_u_var_irf}
  \end{subfigure}

  \caption{IRFs to a standard deviation shock to technology and unemployment in both the estimated state-space model and an unconditional VAR(1) fit to US macroeconomic data for the period 1985-2005.}
  \label{us_zu_irfs}
\end{figure}

\begin{figure}
  \centering
  \begin{subfigure}{0.8\textwidth}
    \includegraphics[width=\linewidth]{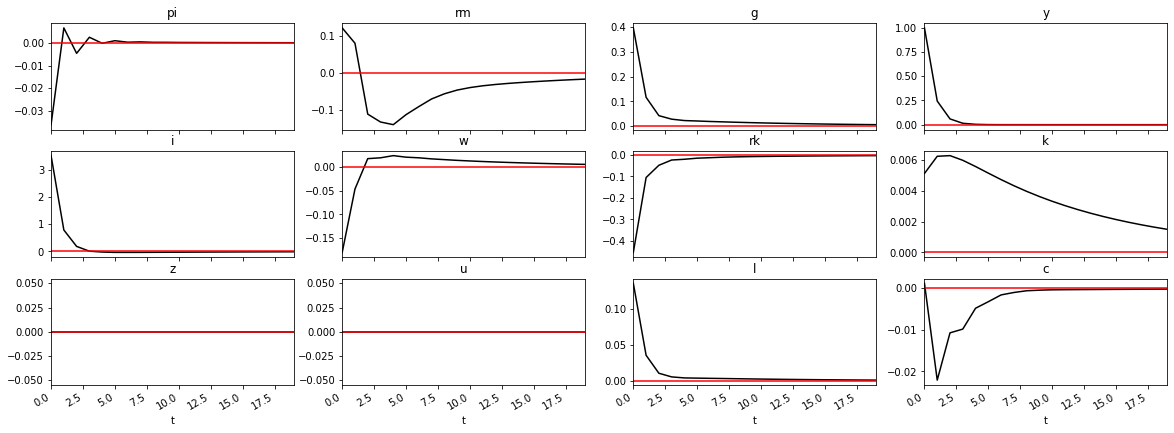}
    \caption{Real Output (State-Space)}
    \label{us_y_irf}
  \end{subfigure}
  \begin{subfigure}{0.8\textwidth}
    \includegraphics[width=\linewidth]{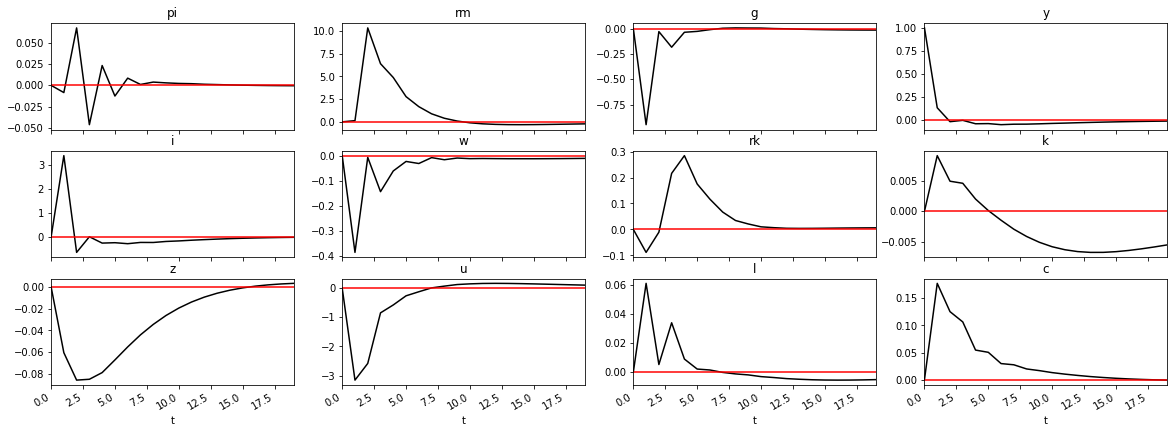}
    \caption{Real Output (VAR)}
    \label{us_y_var_irf}
  \end{subfigure}
  \begin{subfigure}{0.8\textwidth}
    \includegraphics[width=\linewidth]{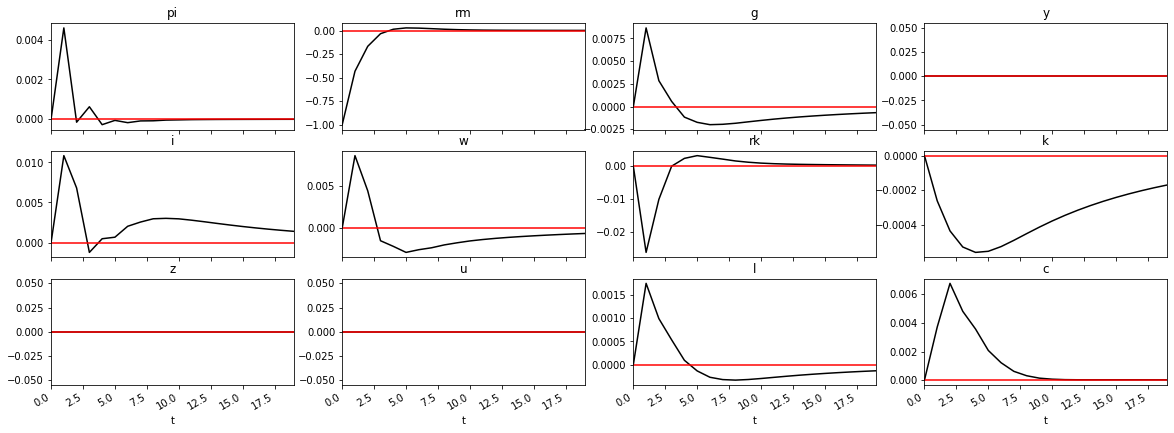}
    \caption{Federal Funds Rate (State-Space)}
    \label{us_rm_irf}
  \end{subfigure}
  \begin{subfigure}{0.8\textwidth}
    \includegraphics[width=\linewidth]{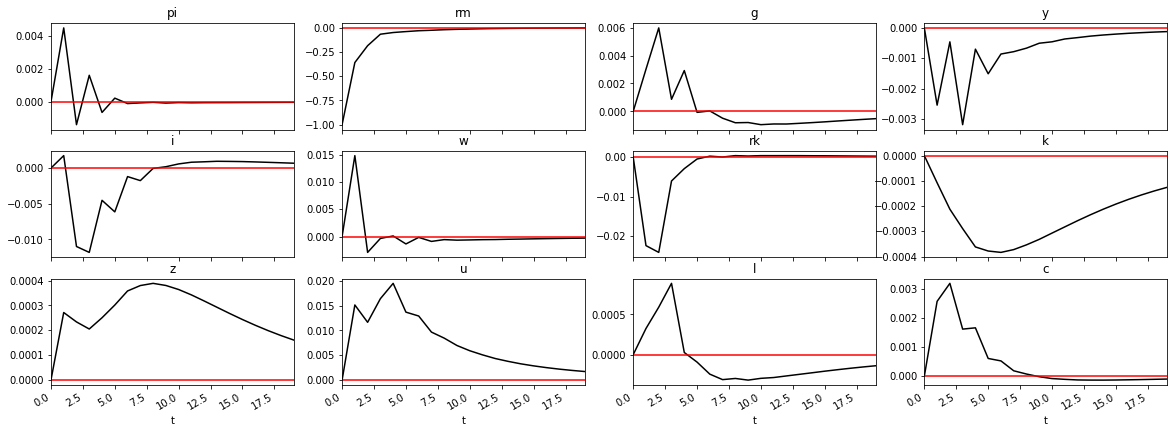}
    \caption{Federal Funds Rate (VAR)}
    \label{us_rm_var_irf}
  \end{subfigure}

  \caption{IRFs to a standard deviation shock to real output and the policy rate in both the estimated state-space model and an unconditional VAR(1) fit to US macroeconomic data for the period 1985-2005.}
  \label{us_yrm_irfs}
\end{figure}

\end{document}